\documentclass[preprint,prd,aps,tighten,nofootinbib,amssymb]{revtex4}
\usepackage{graphicx}
\usepackage{epsfig,amssymb,amsmath}
\usepackage{slashed, color}
\usepackage[makeroom]{cancel}
\usepackage{subfigure}
\usepackage{gensymb}
\usepackage[normalem]{ulem}

\newcommand{\beq}{\begin{equation}}
\newcommand{\eeq}{\end{equation}}
\newcommand{\bea}{\begin{eqnarray}}
\newcommand{\eea}{\end{eqnarray}}
\newcommand{\bear}{\begin{array}}
\newcommand {\eear}{\end{array}}
\newcommand{\bef}{\begin{figure}}
\newcommand {\eef}{\end{figure}}
\newcommand{\bec}{\begin{center}}
\newcommand {\eec}{\end{center}}

\newcommand{\dis}[1]{\begin{equation}\begin{split}#1\end{split}\end{equation}}
\def\vpq{f_{a}}

\newcommand{\lsim}{
\mathrel{\hbox{\rlap{\hbox{\lower4pt\hbox{$\sim$}}}\hbox{$<$}}}}
\newcommand{\gsim}{
\mathrel{\hbox{\rlap{\hbox{\lower4pt\hbox{$\sim$}}}\hbox{$>$}}}}

\newcommand{\C}{\bold c}
\newcommand{\Y}{\tilde{\bold  y}}
\newcommand{\y}{\bold y}

\begin{document}
\draft
\tighten
\preprint{CTPU-17-30}
\preprint{KIAS-P17058}

\title{\large \bf Minimal Flavor Violation with Axion-like Particles}

\author{
    Kiwoon Choi$^{a}$\footnote{email: kchoi@ibs.re.kr},
    Sang Hui Im$^{b}$\footnote{email: shim@th.physik.uni-bonn.de},
    Chan Beom Park$^{c}$\footnote{email: cbpark@kias.re.kr},
    Seokhoon Yun$^{d, a}$\footnote{email: yunsuhak@kaist.ac.kr}}
     
\affiliation{
 $^a$Center for Theoretical Physics of the Universe, \\
 Institute for Basic Science (IBS), Daejeon 34051, Korea \\
 $^b$Bethe Center for Theoretical Physics and Physikalisches Institut der Universit\"at Bonn \\
 Nussallee 12, 53115 Bonn, Germany \\
 $^c$ School of Physics, Korea Institute for Advanced Study, Seoul 02455, Korea\\
 $^d$Department of Physics, KAIST, Daejeon 34141, Korea 
    }

\vspace{4cm}

\begin{abstract}

We revisit the flavor-changing processes involving  an axion-like particle (ALP) in the context of generic ALP effective lagrangian with a discussion of possible UV completions providing the origin of the relevant bare ALP couplings.  
We focus on the minimal scenario that ALP has flavor-conserving couplings at tree level, and the leading flavor-changing couplings  arise  from the loops involving the Yukawa couplings of the Standard Model fermions.
We note that such radiatively generated flavor-changing ALP couplings can be easily suppressed in field theoretic ALP models with sensible UV completion.
We discuss also the implication of our result for string theoretic ALP  originating from higher-dimensional $p$-form gauge fields, for instance for ALP in large volume string compactification scenario.

\end{abstract}

\maketitle

%%%%%%%%%%%%%%%%%%%%%%%%%%
\section{Introduction} 
%%%%%%%%%%%%%%%%%%%%%%%%%%

Axion-like particle (ALP) is a compelling candidate for physics beyond the standard model (BSM) in the intensity frontier searching for a light particle with feeble interactions to the standard model (SM) particles.
Indeed ALP is ubiquitous in many well-motivated BSM scenarios, including the QCD axion introduced to solve the strong CP problem \cite{Peccei:1977hh,Kim:2008hd}, string theories \cite{Choi:1985je,Svrcek:2006yi, Arvanitaki:2009fg},  and the cosmological relaxation of the weak scale \cite{Graham:2015cka}.

Among the various experimental searches for ALP having a mass below few GeV, one of the most sensitive probe is the flavor-changing processes.
Even when the ALP has flavor-conserving couplings to the SM fermions at tree level, there can be  radiatively induced flavor-changing couplings which may yet provide a meaningful constraint on the model \cite{Batell:2009jf, Dolan:2014ska,Freytsis:2009ct}.
In particular, if the ALP has a proper form of tree-level couplings to the SM fermions and/or to the Higgs fields, radiative flavor violation can arise at one-loop, with logarithmically divergent effective couplings proportional to $y^\dagger y \ln \Lambda$, where $y$ denotes the fermion Yukawa couplings and $\Lambda$ is the cutoff scale of the ALP effective theory.   
Such radiatively induced flavor violations have been studied before \cite{Batell:2009jf, Dolan:2014ska}, leading to  a rather strong phenomenological constraint on the model parameters.  
However, these studies are based on the ALP effective interactions which are not manifestly invariant under the electroweak gauge symmetry, while the logarithmic divergence indicates that the dominant contribution comes from the high scales where the electroweak gauge symmetry is restored.
This makes the implication of the previous results \cite{Batell:2009jf, Dolan:2014ska} less clear.

In this paper, we revisit the radiatively induced flavor-changing  ALP couplings in the context of manifestly gauge invariant  effective lagrangian, and examine their implications with a discussion of the possible UV completion providing the origin of the relevant bare ALP couplings.
It is noted that the most dangerous flavor-changing ALP couplings to down-type quarks can be naturally suppressed in field theoretic ALP models with sensible UV completion, in which the ALP originates from the phase degrees of complex scalar fields $X$ charged under the Peccei-Quinn (PQ) symmetry in a UV theory with  cutoff scale significantly higher than the PQ scale $f_a\sim \langle X\rangle$.
The reason is that  bare ALP couplings at $f_a$ are constrained by the condition that the UV theory allows the top quark Yukawa coupling of order unity, which results in a suppression by $1/\tan^2\beta$ of the radiative correction to flavor-changing ALP couplings to down-type quarks, where $\tan\beta =\langle H_2\rangle/\langle H_1\rangle$ for the Higgs doublet $H_2$ responsible for the up-type quark masses and the additional Higgs doublet $H_1$ introduced to accommodate the DFSZ-type ALP in the discussion \cite{Kim:2008hd}.
As a consequence, the flavor constraints on field theoretic ALP  become significantly weaker than the estimation of  \cite{Batell:2009jf, Dolan:2014ska} in the large $\tan\beta$ limit.
Our analysis captures also the result of \cite{Freytsis:2009ct}, which examined the flavor-changing ALP couplings in a model in which the ALP couples to the SM fields through the Higgs bilinear term $H_1H_2$ in the scalar potential of two Higgs doublet model (2HDM).  

We also discuss the  implication of our results for string theoretic  ALP in large volume scenario of string compactification  \cite{Balasubramanian:2005zx}, in which the relevant ALP originates from higher dimensional  $p$-form gauge field with a relatively low  decay constant in phenomenologically interesting range. 
It is noticed that for a given value of $f_a$ set by the couplings to gauge fields,  flavor-conserving tree level couplings of string theoretic ALP to matter fermions are smaller than those of field theoretic ALP by a factor of ${\cal O}(1/16\pi^2)$.
This distinctive feature of stringy ALP  makes the flavor constraints  weaker than the naive expectation, independently of the size of $\tan\beta$.

The organization of this paper is as follows. In the next section, we discuss the radiatively induced flavor-changing ALP couplings in the context of generic ALP effective lagrangian constrained just by the ALP shift symmetry and the SM gauge invariance. 
We consider first the case that the effective theory below the ALP decay constant  is non-supersymmetric, but possibly with additional Higgs doublets, and then discuss the supersymmetric case also.
In Sec.~\ref{sec:UVsupp} , we discuss the possible UV completion of  ALP models, particular the UV origin of the relevant bare ALP couplings.
We  consider two different possibilities, a field theoretic ALP   originating from the phase of PQ-charged complex scalar fields whose vacuum values break the PQ symmetry spontaneously, and a string theoretic  ALP originating from $p$-form gauge field in string theory. 
In Sec.~\ref{sec:Analysis}, we examine the ALP parameter region allowed by phenomenological constraints, including those from the flavor-changing ALP processes, for some specific UV models discussed in Sec.~\ref{sec:UVsupp}.   
Sec.~\ref{sec:conc} is the conclusion.

%%%%%%%%%%%%%%%%%%%%%%%%%%%%%%%%%%%%
\section{Radiatively induced flavor-changing ALP couplings} \label{sec:EFT}
%%%%%%%%%%%%%%%%%%%%%%%%%%%%%%%%%%%%

In this section, we discuss radiatively induced flavor-changing couplings of an axion-like particle in the context of generic effective lagrangian defined at scales above the weak scale, but below the ALP decay constant $f_a$. 
We will use the Georgi-Kaplan-Randall (GKR) field basis   \cite{Georgi:1986df,Choi:1986zw}, in which only the ALP $``a"$ experiences a constant shift, while all other low energy fields $\Phi$ are invariant under the non-linear PQ symmetry:
\bea
U(1)_{\rm PQ}:\,\,\, a\rightarrow a +{\rm constant}, \quad \Phi\rightarrow \Phi.
\eea 
Note that  one can always take such a field basis with an appropriate ALP-dependent field redefinition of the form $\Phi\rightarrow e^{iq_\Phi a/\vpq}\Phi$, where $q_\Phi$ is the PQ charge carried by $\Phi$ in the original field basis. 

In the GKR basis, PQ-invariant ALP interactions at scales below  $\vpq$, which are relevant for our subsequent discussion, can be generally written as\footnote{Here for simplicity we assume the CP invariance, and ignore the terms such as  $\partial_\mu a H_\alpha^Ti\sigma_2 D^\mu H_\beta$ $(\alpha\neq \beta)$ which are assumed to be small in order to forbid the tree level flavor changing neutral current in two (or more) Higgs doublet models. As we consider the effective theory at scales  well above the weak scale, the electroweak gauge symmetry is linearly realized in this ALP effective lagrangian. For a discussion of ALP couplings with non-linearly realized electroweak gauge symmetry, see \cite{Brivio:2017ije}.} 
\dis{
\label{coup_sm}
 {\cal L}_{\rm inv}=\frac{\partial_\mu a}{\vpq}\left[\,\sum_{\psi} (\C_\psi)_{ij} \bar\psi_i \gamma^\mu \psi_j
 +\sum_\alpha c_{H_\alpha}H_\alpha^\dagger\overset{\leftrightarrow}{i D^\mu} H_\alpha\,\right],
}
where $\psi_i=\{Q_i, u^c_i,d^c_i,  L_i, e^c_i\}$ ($i=1,2,3$) stands for the 3 generations of the left-handed quarks and leptons,  and  $H_\alpha$ ($\alpha=1,2$) denote the Higgs doublets with the following Yukawa couplings at scales just below $\vpq$:
\dis{
\label{yukawa_2hdm}
{\cal L}_{\rm Yukawa}=(\Y_u)_{ij} {u^c_i}  Q_{j} H_2 + (\Y_d)_{ij}  {d^c_i} Q_{j} H_d + (\Y_e)_{ij} {e^c_i}  L_j H_e,
}
where each of $H_d$ and $H_e$ can be identified as either $H_1$ or $i \sigma_2 H_2^*$, depending upon the model under consideration.
Making an appropriate ALP-dependent phase rotation of $\psi$ and $H_\alpha$, together with a proper redefinition of the PQ symmetry, one may choose a specific form of GKR  basis for which some of the ALP couplings $(\C_\psi, c_{H_\alpha})$  are vanishing.
However, as we are interested in the UV origin of the above ALP couplings, which will be discussed in the next section, here we take more general field basis which allows a straightforward matching to the UV completion. 
On the other hand,  we limit the discussion to the models with 2HDM, except for the type III 2HDM which can give rise to a tree level flavor changing neutral current (FCNC).
It is straightforward to extend the discussion to models with more Higgs doublets or to the SM without $H_1$, in which $H_d = H_e = i \sigma_2 H_2^*$. 

As $U(1)_{\rm PQ}$ is an approximate symmetry, there can be PQ-breaking ALP interactions also, particularly the non-derivative couplings to gauge fields  and the scalar potential providing a nonzero ALP mass:
\bea
\label{coup_br}
 \Delta {\cal L}_{\rm br}=\frac{a}{\vpq} \sum_{A} C_A \frac{g_A^2}{32 \pi^2} F^{A \mu \nu} \widetilde{F}^A_{\mu \nu}-\frac{1}{2}m_a^2 a^2+...,
 \eea
 where $F^A_{\mu\nu}$ $(A=3,2,1)$ denote the canonically normalized gauge field strength of the SM gauge group $SU(3)_c\times SU(2)_L\times U(1)_Y$.
Here we assume that the ALP mass  is determined by some unspecified UV physics {\it other than} the QCD anomaly, which results in 
\bea
m_a \gg f_\pi m_\pi /f_a.
\eea
Such ALP mass allows that  $f_a$ is small enough to give rise to sizable flavor-changing ALP couplings in the range of phenomenological interest.

We are interested in the case that ALP has flavor-universal couplings to the SM fermions at tree level, so
that the $3\times 3$ ALP coupling matrix $\C_\psi$ takes the flavor-universal form at the cutoff scale $\Lambda_a$ of the ALP effective lagrangian (\ref{coup_sm}): 
\dis{
\label{MF}
(\C_\psi)_{ij} (\mu =\Lambda_a) = c_\psi\, \delta_{ij} \quad \big( \psi=\{Q, u^c, d^c, L, e^c\}\big)\, .
} 
For field theoretic ALP originating from the phase of PQ charged complex scalar fields, $\Lambda_a$ can be identified as the scale where the PQ symmetry is spontaneously broken, i.e. $$\Lambda_a\sim f_a.$$
On the other hand, for string theoretic ALP from higher-dimensional $p$-form gauge field, one finds \cite{Choi:1985je,Svrcek:2006yi, Arvanitaki:2009fg} $$\Lambda_a\sim M_{\rm st} \sim \frac{8\pi^2}{g^2} f_a,$$ where $M_{\rm st}$ is the string scale and the additional  factor $8\pi^2/g^2$ originates from the convention to define the ALP decay constant through the ALP interaction to gauge fields in (\ref{coup_br}), while assuming  $C_A={\cal O}(1)$.

Even when $\C_\psi$ are flavor-universal at $\Lambda_a$,  non-universal piece can be generated by radiative corrections at lower scales.
The leading part of those radiative corrections can be captured by the following form of one-loop renormalization group (RG) equations, which can be determined  up to an overall coefficient $\xi$ by the covariance under the $SU(3)$ flavor rotations of $\psi$ and the ALP-dependent field redefinitions $\psi\rightarrow e^{ix_\psi a/f_a} \psi, \, H_\alpha\rightarrow e^{ix_\alpha a/f_a}H_\alpha$: 
\bea
\frac{d\C_Q}{d\ln \mu} & = & \frac{\xi}{32 \pi^2}\left( \, \C_Q \big(\Y_u^\dagger  \Y_u + \Y_d^\dagger \Y_d\big)
 + \Y_u^\dagger \C_{u^c}^T   \Y_u + \Y_d^\dagger \C_{d^c}^T \Y_d \right.\nonumber \\
 && \left. \quad \quad \quad  +\,c_{H_2}\Y_u^\dagger \Y_u +c_{H_d}\Y_d^\dagger \Y_d 
 +{\rm h.c}\right),  \nonumber \\
\frac{d\C_{u^c}^T}{d\ln \mu} & = & \frac{\xi}{16\pi^2}\left( \, \Y_{u} \C_Q \Y_{u}^\dagger +  \C_{u^c}^T \Y_{u} \Y_{u}^\dagger + c_{H_2}\Y_u \Y_u^\dagger + {\rm h.c}  \right),  \nonumber \\
\frac{d\C_{d^c}^T}{d\ln \mu} & = & \frac{\xi}{16\pi^2}\left( \, \Y_{d} \C_Q \Y_{d}^\dagger +  \C_{d^c}^T \Y_{d} \Y_{d}^\dagger + c_{H_d}\Y_d \Y_d^\dagger + {\rm h.c}  \right), \nonumber  \\
\frac{d\C_L}{d\ln \mu} & = & \frac{\xi}{32\pi^2}\left( \, \C_L \Y_e^\dagger \Y_e +\Y_e^\dagger \C_{e^c}^T \Y_e  +c_{H_e}\Y_e^\dagger \Y_e+{\rm h.c} \right), \nonumber  \\
\frac{d\C_{e^c}^T}{d\ln \mu} & = & \frac{\xi}{16\pi^2}\left( \, \Y_{e} \C_L \Y_{e}^\dagger +  \C_{e^c}^T \Y_{e} \Y_{e}^\dagger + c_{H_e}\Y_e \Y_e^\dagger +{\rm h.c} \right), \label{RG}
\eea
where we include {\it only} the Yukawa-dependent parts which can  generate  flavor-changing ALP couplings at low energy scales. 
In the following, we will use the above RG equation at the leading log approximation to derive the ALP-fermion coupling $\C_\psi$ around the weak scale.
Note that the radiatively generated flavor-changing ALP couplings are produced dominantly  by the loops involving the top quark and the (Goldstone-mode) Higgs fields, which would be encoded in the RG running from $f_a$ down to the weak scale.
\begin{figure} 
\centering
\includegraphics[scale=0.43]{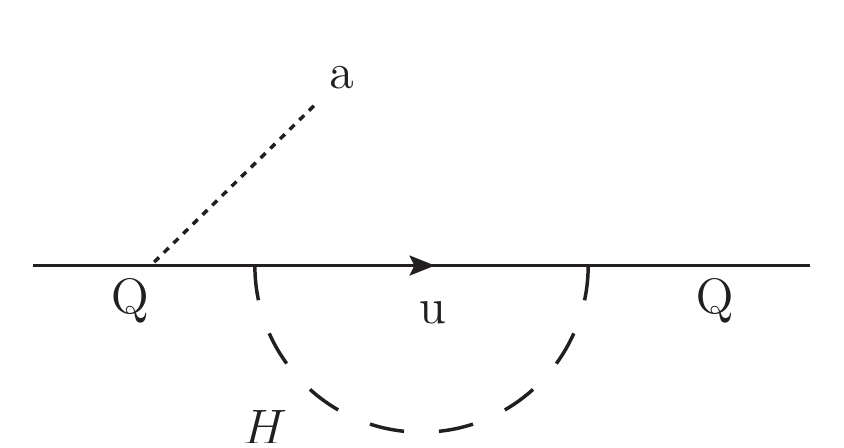}
\includegraphics[scale=0.43]{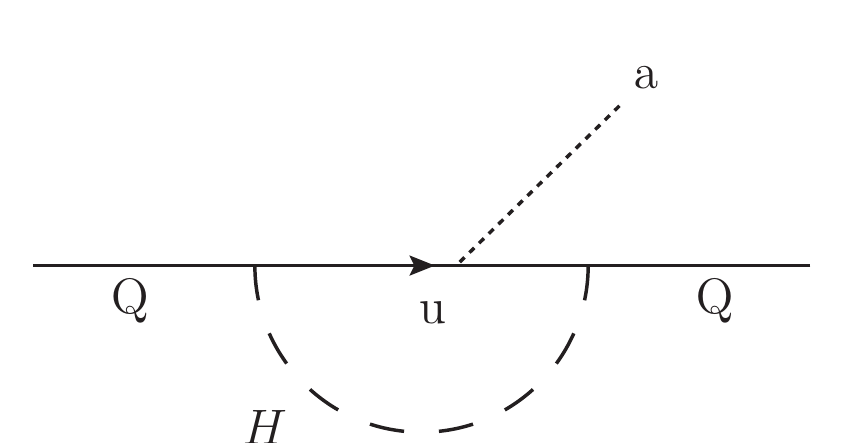}
\includegraphics[scale=0.43]{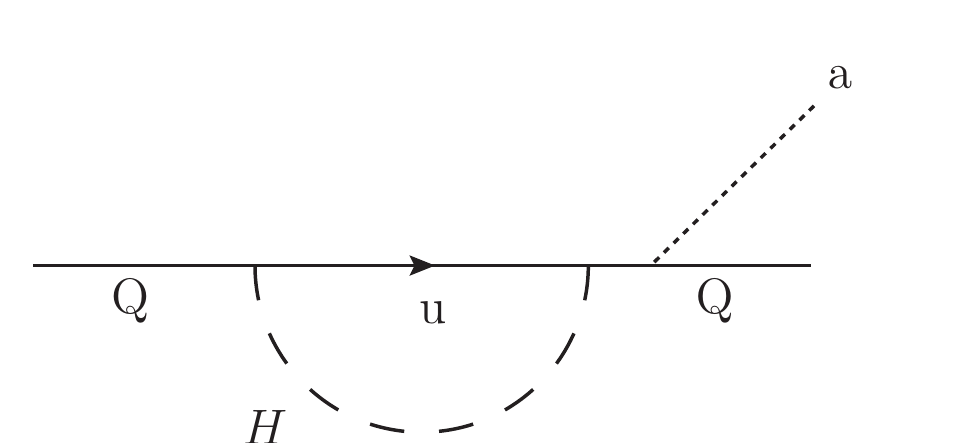} \hspace{-0.7cm}
\includegraphics[scale=0.43]{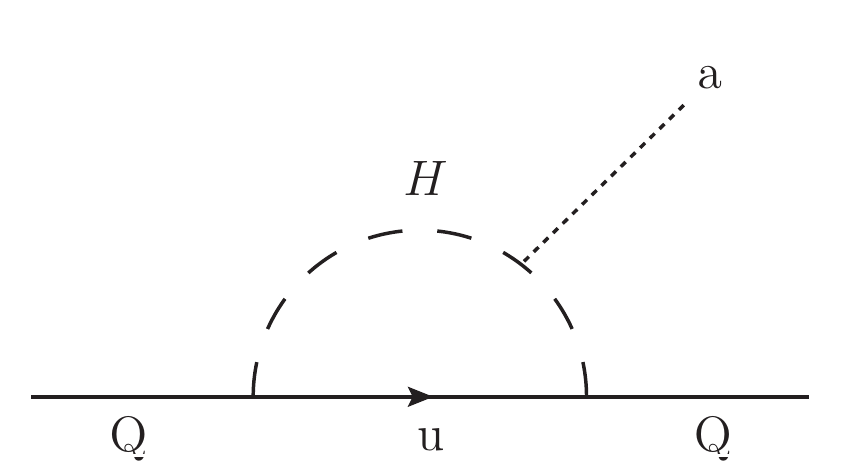}
\caption{One-loop diagrams for the running of $\C_\psi$}
\label{fig:running_cpsi}
\end{figure}
At any rate, for non-supersymmetric ALP model, one easily finds from the diagrams in Fig. \ref{fig:running_cpsi} that the RG coefficient $\xi$  is given by
\bea
\big(\xi\big)_{\rm non-SUSY}=1.
\eea

In supersymmetric (SUSY) ALP models, there can be additional diagrams involving the superpartner particles, which would contribute to the RG coefficient $\xi$ in (\ref{RG}).
One the other hand, in SUSY models there is a simple connection  between the beta function of ALP coupling  and the anomalous dimension of chiral matter field \cite{ArkaniHamed:1998kj}, with which one can easily compute the RG coefficient  $\xi$.
To see this, we first note that in SUSY model, the ALP interaction (\ref{coup_sm}) can be encoded in the following superfield interactions
\bea
\int d^4\theta \, (c_\Phi)_{IJ}\frac{(A+A^*)}{\vpq}\Phi^{*I}\Phi_J 
\eea
where $\Phi_I$ denote the chiral superfields including the quark and lepton superfields, as well as the Higgs doublet superfields in SUSY models, and $A$ is the ALP superfield which contains the saxion ($s$) and the axino ($\tilde a$) as
$$A= (s+ia)+\sqrt{2}\theta\tilde a + \theta^2 F^A.
$$ 
To proceed, it is enough to consider a toy  model involving the ALP superfield and a single chiral matter superfield $\Phi$, with the following effective lagrangian
\bea
\int d^4\theta \, Z_\Phi\Phi^*\Phi +\left(\int d^2\theta\,
\frac{1}{3}\lambda_\Phi \Phi^3+{\rm h.c}\right), 
\eea
where 
\bea
Z_\Phi=Z_0\left(1 +c_\Phi\frac{(A+A^*)}{\vpq}\right)\, ,
\eea
and  $Z_0$ and $\lambda_\Phi$ are constants.  
One then finds 
\dis{
\label{c_susy}
\bold c_\Phi = \vpq \left. \frac{\partial \ln Z_\Phi}{\partial A} \right|_{A=0}, 
}
and therefore 
\dis{
\label{cr_susy}
\frac{d \, \bold c_\Phi}{d \ln \mu} = \vpq \left. \frac{\partial}{\partial A} \left( \frac{d \ln Z_\Phi}{d \ln \mu} \right) \right|_{A=0}.
}
On the other hand, $d\ln Z_\Phi/d\ln\mu$ corresponds to the superspace anomalous dimension, whose one-loop expression is given by 
\bea
\frac{d \ln Z_\Phi}{d \ln \mu} =-\frac{1}{8\pi^2}\frac{\lambda_\Phi^*\lambda_\Phi}{Z_\Phi Z_\Phi Z_\Phi}.
\eea
It is straightforward to generalize this observation to the ALP couplings to the MSSM chiral superfields,  from which we find that the RG coefficient $\xi$ in SUSY ALP model is given by
\bea
\big(\xi\big)_{\rm SUSY}=2,
\eea
with $c_{H_d} = c_{H_e} = c_{H_1}$.
Note that here we consider only the minimal radiative flavor violation induced by the Yukawa couplings of the SM fermions, while ignoring other sources of flavor violation which might exist in SUSY models.

The BSM degrees of freedom in our ALP model, i.e. additional Higgs doublet and/or the superpartners, might have a mass well above the weak scale. In such case, we should integrate out those BSM particles to derive the ALP couplings at the weak scale.  
For simplicity, we assume that all BSM particles have a similar mass $m_{\rm BSM}$ which would correspond to the charged Higgs boson mass in the 2HDM, $m_{\rm BSM}=m_{H^\pm}$, or the superpartner masses in SUSY ALP models,  $m_{\rm BSM}=m_{\rm SUSY}$.

In the process to integrate out the BSM particles at $m_{\rm BSM}\gg m_W$, the only matching condition relevant for low energy ALP couplings in our approximation is those for the Higgs doublets, which are given by 
\bea
H_1= H^*\cos\beta, \quad H_2=H\sin\beta \, ,
\label{matching_cond}
\eea
where $H$ corresponds to the SM Higgs doublet.
Then the PQ invariant ALP couplings at scales below $m_{\rm BSM}$ are given by
\dis{
\label{coup}
{\cal L}_{\rm inv}=\frac{\partial_\mu a}{\vpq}\left[\,\sum_{\psi} (\C_\psi)_{ij} \bar\psi_i \gamma^\mu \psi_j
+ c_{H}H^\dagger\overset{\leftrightarrow}{i D^\mu} H\,\right],
}
with the matching condition
\bea \label{cH_BSM}
c_H(\mu=m_{\rm BSM})=c_{H_2}\sin^2\beta - c_{H_1}\cos^2\beta \, ,
\eea
and the SM Yukawa couplings
\dis{
\label{yukawa_sm}
{\cal L}_{\rm Yukawa}=(\y_u)_{ij} {u^c}_{i}  Q_{j} H + (\y_d)_{ij}  {d^c}_{i} Q_{j} H^* + (\y_e)_{ij} {e^c}_{i}  L_j H^*\, ,
}
where
\bea
\y_u = \Y_u\sin\beta, \quad \y_d =\Y_d
\cos \beta~\, \textrm{or}\, ~\Y_d\sin\beta, \quad \y_e =\Y_e\cos\beta \, ~\textrm{or}\,~ \Y_e\sin \beta,
\eea
where the matching conditions for $\y_d$ and $\y_e$ depend on the type of 2HDM under  consideration. 
The relevant RG evolution of ALP couplings from $m_{\rm BSM}$ to the weak scale are given by
\bea
\frac{d\C_Q}{d\ln \mu} & = & \frac{1}{32 \pi^2}\left(\C_Q \big(\y_u^\dagger  \y_u + \y_d^\dagger \y_d\big)
 + \y_u^\dagger \C_{u^c}^T   \y_u + \y_d^\dagger \C_{d^c}^T \y_d \right.\nonumber \\
 &&\left. \quad \quad \quad  +\, c_{H}\big(\y_u^\dagger \y_u -\y_d^\dagger \y_d\big) 
 +{\rm h.c}\right),  \nonumber \\
\frac{d\C_{u^c}^T}{d\ln \mu} & = & \frac{1}{16\pi^2}\left(  \, \y_{u} \C_Q \y_{u}^\dagger +  \C_{u^c}^T \y_{u} \y_{u}^\dagger + c_{H}\y_u \y_u^\dagger + {\rm h.c}  \right),  \nonumber \\
\frac{d\C_{d^c}^T}{d\ln \mu} & = & \frac{1}{16\pi^2}\left(  \, \y_{d} \C_Q \y_{d}^\dagger +  \C_{d^c}^T \y_{d} \y_{d}^\dagger -c_{H}\y_d \y_d^\dagger + {\rm h.c}  \right), \nonumber  \\
\frac{d\C_L}{d\ln \mu} & = & \frac{1}{32\pi^2}\left( \, \C_L \y_e^\dagger \y_e +\y_e^\dagger \C_{e^c}^T \y_e  -c_{H}\y_e^\dagger \y_e+{\rm h.c} \right), \nonumber  \\
\frac{d\C_{e^c}^T}{d\ln \mu} & = & \frac{1}{16\pi^2}\left(\, \y_{e} \C_L \y_{e}^\dagger +  \C_{e^c}^T \y_{e} \y_{e}^\dagger -
c_{H}\y_e \y_e^\dagger +{\rm h.c} \right). \label{RG_SM}
\eea

The RG induced non-universal elements of $\C_\psi$ will lead to flavor-changing ALP interactions at low energy scales after rotating to the fermion mass eigenbasis.
The dominant experimental constraints on flavor-changing ALP interactions come from the down-type quark processes. In the mass eigenbasis, the ALP couplings to the left-handed down-type quarks are given by
\dis{
 c^d_{ij}\frac{\partial_\mu a}{\vpq}   \bar{d}_{L i} \gamma^\mu  {d_L}_j ~\rightarrow~ -i  c^d_{ij} \frac{a}{\vpq}
 \bar{d}_i \left(m_{d_i} P_L - m_{d_j} P_R \right) d_j,
}
where 
\bea
c^d_{ij}=(U_{d_L}^\dagger \C_Q U_{d_L})_{ij},
\eea
and $d_{L i} \, (d_i)$ denote the left-handed (Dirac) down-type quark fields in the mass eigenbasis, which is obtained by the unitary rotation $d_{L} \rightarrow U_{d_L} d_{L}.$
Here we used the equations of motion of the fermion fields to get the last expression.   
Applying the one-loop RG equations (\ref{RG}) and (\ref{RG_SM}), we find
\bea
\label{fcnc_down}
c^d_{ij}&=&  -\frac{\xi}{16\pi^2} \left(c_Q+c_{u^c} +c_{H_2}\right)\left(V_{\rm CKM}^\dagger \, \Y_{u}^{D \dagger} \Y_{u}^D \, V_{\rm CKM} \right)_{ij} \ln\left(\frac{\Lambda_a}{m_{\rm BSM}}\right) \nonumber \\
&&-\frac{1}{16\pi^2} \left(c_Q+c_{u^c} +c_{H}\right)\left(V_{\rm CKM}^\dagger \, \y_{u}^{D \dagger} \y_{u}^D \, V_{\rm CKM}  \right)_{ij} \ln\left(\frac{m_{\rm BSM}}{\mu}\right) + \dots,
\nonumber \\ 
&\approx&  -\frac{m_t^2}{16\pi^2 v^2} (V_{\rm CKM})_{3i}^* (V_{\rm CKM})_{3j}
\left[\frac{\xi}{\sin^2\beta} \left(c_Q+c_{u^c}+c_{H_2}\right)
 \ln\left(\frac{\Lambda_a}{m_{\rm BSM}}\right)\right.\nonumber \\
 &&\left.+\, \left(c_Q+c_{u^c}+c_{H_2}-(c_{H_1}+c_{H_2})\cos^2\beta\right)\ln\left(\frac{m_{\rm BSM}}{m_t}\right) 
 \right] + \dots,
 \label{downTypeQuarkALPcoup}
\eea
where $\y_\psi^D$ denotes the diagonalized Yukawa matrices in the CKM basis, $v= 174$ GeV, and the ellipses stand for the irrelevant flavor-diagonal parts.
Note  that the down-type Yukawa couplings do not give rise to a flavor-violating coupling of the down-type quarks at one-loop approximation due to the GIM mechanism.
Likewise, the other ALP couplings  $\C_\psi \, (\psi = u^c, d^c, e^c, L)$ in the one-loop approximation  are diagonalized in the CKM basis as long as the flavor-universal condition (\ref{MF}) is satisfied at the scale $\vpq$, so they do not generate a flavor violation at one-loop.\footnote{If we include the right-handed neutrinos with proper Yukawa couplings, the one-loop corrected ALP-lepton coupling $\C_L$ includes flavor-changing piece proportional to the square of the right-handed neutrino Yukawa couplings.
 However, such lepton-flavor-changing ALP couplings can be safely ignored because either the right-handed neutrinos are superheavy or the neutrino Yukawa couplings are negligibly  small in order to be compatible with the observed small neutrino masses.}

From (\ref{fcnc_down}), we find that in the large $\tan\beta$ limit, the flavor-changing processes like $b \rightarrow s+ a$ or $s \rightarrow d+ a$  can happen with a sizable rate {\it if} $c_Q + c_{u^c} + c_{H_2}$  has a non-zero value. 
In case that $c_Q + c_{u^c} + c_{H_2}=0$, the next leading order contribution arises from a non-zero value of $c_{H_1} + c_{H_2}$, multiplied by an additional suppression factor $1/\tan^2 \beta$.
In the next section, we will discuss the implication of this point in terms of the possible UV completion of the ALP effective coupling (\ref{coup_sm}).
Especially, we will see that this  implies a suppression of the flavor-changing ALP couplings to the down-type quarks for field theoretic ALP with a sensible UV completion.

We emphasize that the above expression (\ref{fcnc_down}) of the ALP coupling to the down-type quarks  is independent of the type of 2HDM under consideration, as far as the SM-like Higgs in the decoupling limit, i.e. $H_2$ in our convention,   couples only to the up-type quark sector, which is the case for all 2HDMs not involving FCNC at tree-level.
Furthermore, even in models involving more Higgs doublets beyond the 2HDMs, the suppression factor $1/\tan^2 \beta$ should generically appear.  
This can be shown by considering the corresponding generalization of the matching condition (\ref{cH_BSM}) for multiple Higgs doublet models.
If we define $\sin \beta \equiv \langle H_2 \rangle/\langle H \rangle$ (i.e. the ratio of the vacuum value of the Higgs of the up-type quark sector to the SM Higgs vacuum value), 
the matching condition is generalized to
\dis{
 c_H(\mu=m_{\rm BSM})=c_{H_2}\sin^2\beta + \cos^2\beta \left( \sum_{\alpha \neq 2} 2 Y_{H_\alpha} c_{H_\alpha} \frac{v_\alpha^2}{\sum_{\beta \neq 2} v_\beta^2}  \right) \, ,
 }
where $v_{\alpha} \equiv \langle H_\alpha \rangle$, and $Y_{H_\alpha}$ denotes the $U(1)_Y$ hypercharge of $H_{\alpha}$ which should be either 1/2 or -1/2 to preserve the electromagnetic
$U(1)_{\rm EM}$ symmetry. 
Then
\dis{
c_{Q} + c_{u^c} + c_H = c_{Q} + c_{u^c} + c_{H_2} - \left[ c_{H_2} - \left( \sum_{\alpha \neq 2} 2 Y_{H_\alpha} c_{H_\alpha} \frac{v_\alpha^2}{\sum_{\beta \neq 2} v_\beta^2}  \right)
\right] \cos^2 \beta.
}
Therefore, if $c_{Q} + c_{u^c} + c_{H_2} = 0$, the dominant term in (\ref{fcnc_down}) is still accompanied by $1/\tan^2 \beta$.
We also comment that inclusion of singlet fields or $SU(2)_L$-triplet fields etc in the Higgs sector contributes to the flavor violation only by higher dimensional operators and does not change our results at leading order.

Flavor-changing  ALP couplings to the up-type quarks can be similarly derived from the one-loop corrected $\C_Q$. 
Contrary to the case of down-type quarks, the couplings to the up-type quarks  depend   on the type of 2HDM under consideration.
In the mass eigenbasis, the resultant  ALP couplings turn out to be
\dis{ 
-i \frac{a}{\vpq}\, c^u_{ij}\, \bar{u}_i \left(m_{u_i} P_L - m_{u_j} P_R \right) u_j 
\label{eq:upFCNCoperator}
}
with
 \bea
\label{fcnc_up}
c^u_{ij} &=&  -\frac{\xi}{16\pi^2}  \left(c_Q+c_{d^c} +c_{H_d}\right)\left(V_{\rm CKM}\,  \Y_{d}^{D \dagger} \Y_{d}^D \,V_{\rm CKM}^\dagger \right)_{ij} \ln\left(\frac{\Lambda_a}{m_{\rm BSM}}\right) \nonumber \\
&&-\frac{1}{16\pi^2} \left(c_Q+c_{d^c} -c_{H}\right)\left(V_{\rm CKM} \, \y_{d}^{D \dagger} \y_{d}^D \, V_{\rm CKM}^\dagger  \right)_{ij} \ln\left(\frac{m_{\rm BSM}}{\mu}\right) + \dots,
\nonumber \\ 
&\approx&  -\frac{m_b^2}{16\pi^2 v^2} (V_{\rm CKM})_{i 3} (V_{\rm CKM})_{j 3}^* 
\left[\xi \left(c_Q+c_{d^c}+c_{H_d}\right) \left\{ \begin{array}{l}
1/\cos^2 \beta \\
1/\sin^2 \beta
\end{array}\right\}
 \ln\left(\frac{\Lambda_a}{m_{\rm BSM}}\right)\right.\nonumber \\
 &&\left.+\, \left(c_Q+c_{d^c}+c_{H_d}-(c_{H_1}+c_{H_2})\left\{ \begin{array}{l}
\sin^2 \beta \\
-\cos^2 \beta
\end{array}\right\}\right)\ln\left(\frac{m_{\rm BSM}}{m_W}\right) 
 \right] + \dots,
 \eea
where the upper entry of the column applies to the SUSY, type II and type Y 2HDMs with $c_{H_d} = c_{H_1}$, while the lower entry corresponds to the type I and type X 2HDMs with $c_{H_d} = -c_{H_2}$, and the ellipses denote the flavor-diagonal part.
Here we see that flavor-changing ALP couplings to the up-type quarks arise from the down-type Yukawa couplings, while the up-type Yukawa couplings generate only a flavor-conserving piece due to the GIM mechanism, and therefore the resultant couplings are suppressed by small $m_b^2/m_t^2$ compared to the couplings to the down-type quarks. 
Moreover, the experimental sensitivity of the up-type quark sector to a new physics involving FCNC process is known to be rather weak as it is screened by the QCD long distance effect \cite{Burdman:2001tf}.    
Yet, in certain models such as SUSY,  type II  and type Y 2HDMs, the couplings are multiplied by $\tan^2 \beta$, and therefore can be sizable in the large  $\tan \beta$ limit. This is because the $b$-quark Yukawa coupling is enhanced by $\tan \beta$ at scales above the BSM scale $m_{\rm BSM}$.  
As a result,  depending on the type of 2HDM under consideration, the flavor-changing processes of the up-type quarks  might impose a meaningful constraint on the ALP decay constant $\vpq$. In the next section, we will address this point  with a discussion of possible UV completion of ALP models.
 
%%%%%%%%%%%%%%%%%%%%%%%%%%%%%%%%%%%%
\section{implication for UV completed ALP models} \label{sec:UVsupp}
%%%%%%%%%%%%%%%%%%%%%%%%%%%%%%%%%%%%

In this section, we discuss possible UV completion of ALP models to examine the implication of radiatively induced flavor-changing ALP interactions. 
As for the UV origin of ALP, there are two possibilities. ALP might originate from the phase of PQ-charged complex scalar fields whose vacuum values break the PQ symmetry  spontaneously, which we call  field theoretic ALP, or from higher dimensional $p$-form gauge fields in UV theory with extra spacial dimension, which we call string theoretic ALP.
For both type of ALPs, there exist a scalar partner in the UV theory, i.e. the radial mode of PQ-breaking complex scalar field for field theoretic ALP  and the modulus partner of string theoretic ALP, whose vacuum value determines the ALP decay constant $\vpq$.
As we will see, the ALP couplings to the SM fermions and the Higgs doublets, which are of our primary concern,  have a definite connection to the couplings of the scalar partner in the Yukawa sector and the Higgs potential.

\subsection{Field theoretic ALP}

Let us first consider an ALP originating from the phase of PQ-charged complex scalar fields.
For simplicity, we assume  that the ALP corresponds mostly to the phase of a single complex scalar field $X$ with PQ charge $q_X=-1$:
\bea
X = \frac{1}{\sqrt{2}}\, \rho\, e^{ia/\vpq},
\eea
where the vacuum value of the radial mode,  $\langle\rho\rangle=\vpq$, can be identified as the ALP decay constant in low energy effective theory.
Generically this PQ-charged $X$ can couple to the Yukawa sector and the Higgs potential as 
\bea
\label{yukawa_uv}
\left(\frac{X}{M_*}\right)^{q_{u_i}+q_{Q_j}+q_{H_2}}(\lambda_u)_{ij}{u}^c_{i}  Q_{j} H_2 +  \left(\frac{X}{M_*}\right)^{q_{d_i}+q_{Q_j}+q_{H_d}} (\lambda_d)_{ij} {d}^c_{i} Q_{j} H_d \nonumber \\
 +  \left(\frac{X}{M_*}\right)^{q_{e_i}+q_{L_j}+q_{H_e}} (\lambda_e)_{ij}{e}^c_{i}  L_j H_e  +
  b_0 \left(\frac{X}{M_*}\right)^{q_{H_1}+q_{H_2}} H_1 H_2 + {\rm h.c.},
\eea
where $q_I$ denote the PQ charge of the corresponding field $\Phi_I$, $M_*$ is the cut-off scale of the above effective interactions, which should be bigger than $f_a$ for consistency, and $b_0$ is a  parameter with mass-dimension two. 
Again we remark that each of $H_d$ and $H_e$ corresponds to either $H_1$ or $i \sigma_2 H_2^*$ depending on the type of 2HDM under consideration.

After replacing $X$ with its vacuum value,
\bea
\langle X\rangle = \frac{1}{\sqrt{2}} f_a e^{ia/f_a},
\eea
the UV Yukawa couplings in (\ref{yukawa_uv}) can be matched to the effective theory Yukawa couplings (\ref{yukawa_2hdm}) in the GKR field basis, with an ALP-dependent field redefinition
\bea
\Phi_I \rightarrow e^{-iq_I a/\vpq} \Phi_I \quad \big(\Phi_I=\psi_i, H_{1,2}\big),
\eea
which results in the following matching condition\footnote{Note that there can be a small correction of ${\cal O}(\vpq^2/M_*^2)$ to this matching condition due to the higher-dimensional operators such as $\frac{X^*\partial_\mu X}{M_*^2}\bar\psi\gamma^\mu\psi$, which will be ignored in the following discussion.} for the ALP couplings at the scale $f_a$:
\bea \label{matching_1}
\big(\C_\Phi\big)_{IJ} (\mu=\vpq) = q_I\delta_{IJ}.
\eea
It is an interesting possibility that the PQ charges are flavor-non-universal in such a way that the observed hierarchical masses and mixing angles of charge fermion originate from the PQ-breaking spurion factor $(X/M_*)^{q_{\psi_i}+q_{\psi_j}+q_H}$ \cite{Ema:2016ops, Calibbi:2016hwq}.  
However, in such case ALP has flavor-changing couplings at tree-level, and the radiative corrections discussed in the previous section give only a small subleading correction to the tree level result.

If the PQ charges of the SM fermions are flavor-universal, i.e.
\bea
q_{\psi_i}=q_\psi \quad (\psi=Q,u^c,d^c,L,e^c),
\label{eq:FlavorUniversalCoupling}
\eea
then there is no flavor-changing ALP coupling at tree level, and the one-loop radiative corrections discussed in the previous section might provide the dominant source of  flavor violating ALP processes at low energy scales.
After the spontaneous breaking of PQ symmetry, the fermion Yukawa couplings  and the coefficient of the Higgs bilinear term are given by 
\bea
(\Y_\psi)_{ij} &=& \left(\frac{\vpq}{M_*}\right)^{n_\psi}(\lambda_\psi)_{ij}  \quad \big(\psi=u,d,e\big), \\
\label{eq:SpurionYukawa}
{b} &=& \left(\frac{\vpq}{M_*}\right)^{n_H} b_0,
\eea
where the non-negative integer $n_\psi$ and $n_H$ are given by
\bea \label{n_s}
n_u&=& q_{Q}+q_{u^c}+q_{H_2}= c_{Q} + c_{u^c} +c_{H_2}, \nonumber \\
n_d&=& q_{Q}+q_{d^c}+q_{H_d}= c_{Q} + c_{d^c} +c_{H_d},\nonumber \\
n_e&=& q_{L}+q_{e^c}+q_{H_e}= c_{L} + c_{e^c} +c_{H_e},\nonumber \\
n_H&=& q_{H_1} + q_{H_2} = c_{H_1} + c_{H_2}.
\eea
One then  finds from (\ref{fcnc_down}) and (\ref{fcnc_up}) that $n_{u,d}$  correspond to the coefficients of RG running   generating the flavor-changing ALP couplings starting from the ALP scale $\vpq$.
In other words,  a {\it nonzero} value of $n_{u,d}$  can be identified as the dominant source of flavor-changing ALP couplings, which would be enhanced by the large logarithmic factor $\ln (\vpq/m_{t,W})$.
Also one finds that $n_H$ corresponds to the RG running coefficient generating flavor-changing ALP couplings starting from the BSM scale $m_{\rm BSM}$  to the weak scale.

Obviously   it is not possible to get the correct top quark Yukawa coupling with {\it nonzero} $n_u$, while satisfying the perturbativity bound $\lambda_t \lesssim {\cal O}(1)$, unless the cutoff scale $M_*$ is comparable to $\vpq$. 
Although $M_*$ can be  determined only by the next step of UV completion, which is beyond the scope of this work, there is neither theoretical nor phenomenological motivation for $M_*\sim f_a$.
Rather, the spontaneous breaking of PQ symmetry should be interpreted as an IR phenomenon even within the present level of UV completion, which means that it is implicitly assumed that the cutoff scale $M_*\gg f_a$.
As we will see in the next section, the radiatively generated flavor-changing ALP couplings discussed in the previous section can be sizable enough to be phenomenologically relevant, {\it only when} $f_a\lesssim 10^7$ GeV. 
For such low PQ scale, the cutoff scale $M_*$ of the present level of UV completion involving the effective interaction (\ref{yukawa_uv}) is likely to be much higher than $f_a$,  for instance  at least by one order of magnitude.
This implies that
\bea
n_u=c_Q+c_{u^c}+c_{H_2}=0
\eea
for generic field theoretic ALP which has a sensible UV completion.
Then the flavor-changing ALP couplings to the down-type quarks  start to be radiatively generated only from the scale $m_{\rm BSM}$, with a further suppression by $1/\tan^2\beta$ (see eq. (\ref{fcnc_down})).
In fact, the RG-induced flavor-changing ALP couplings generated at scales below $m_{\rm BSM}$ correspond to the leading piece  of the finite result calculated in \cite{Freytsis:2009ct} for a specific UV-completed ALP model in which the ALP couples to the SM sector only through the Higgs bilinear term $H_1 H_2$,  which amounts to the case with $n_H \neq 0$ and $n_u=n_d =n_e=0\,$ in our terminology. 
Our discussion suggests  that the suppression by $1/\tan^2 \beta$  of the one-loop flavor violation is rather generic, and applies for a wide class of field theoretic ALP models beyond the specific example discussed in \cite{Freytsis:2009ct}.  

Since the major constraint on the ALP decay constant $\vpq$ comes from  the down-type quark sector, the above observation suggests that the constraints on the ALP models with $n_u = 0$ will be significantly weaker than the previous results which have been obtained  based on a simple ansatz for the tree level ALP couplings \cite{Batell:2009jf, Dolan:2014ska}, which is in fact hard to be realized within a sensible field theoretic UV completion.
For instance, the Yukawa-like ALP couplings\footnote{If non-derivative ALP couplings are used for the calculation as in \cite{Dolan:2014ska}, one must take into account the additional couplings  $i (\tilde{y}_u)_{ij} \frac{a}{\vpq} u_{R i}^c d_{L j} H_2^+ + i (\tilde{y}_d)_{ij} \frac{a}{\vpq} d_{R i}^c u_{L j} H_d^-$  in order to maintain the gauge invariance.
Similarly, if one considers  an ALP coupling to quark axial vector  current as in \cite{Batell:2009jf}, the associated coupling to vector current must be included for the gauge invariance.
This additional vector current coupling can be rotated to the couplings  $i (\tilde{y}_u)_{ij} \frac{a}{\vpq} u_{R i}^c d_{L j} H_2^+ + i (\tilde{y}_d)_{ij} \frac{a}{\vpq} d_{R i}^c u_{L j} H_d^-$ by an appropriate ALP-dependent redefinition of the quark fields.} assumed  in \cite{Dolan:2014ska} correspond to the case of $n_u = n_d \neq 0$, which can not be achieved from field theoretic UV completion with a cutoff scale $M_*$ significantly higher than $f_a$. 
The universal ALP couplings assumed in \cite{Batell:2009jf} correspond to the case of $ q_Q=q_{u^c} = q_{d^c} \neq 0$ and $q_{H_1} = q_{H_2} = 0$, which again can not be achieved from sensible field theoretic UV completion.

Given that $n_u=0$ for field theoretic ALP models with sensible UV completion, and as a result the one-loop flavor-changing ALP couplings to the down-type quarks  are suppressed by $1/\tan^2 \beta$, higher loop effects might be even more important than the one-loop contribution if $\tan \beta$ is large enough. 
Recently, it was pointed out that the following ALP coupling to  the $W$-bosons, 
\dis{
\label{aWW}
C_{aWW} \frac{a}{\vpq} \frac{g_2^2}{32\pi^2} W \widetilde{W},
}
which might exist as a part of  (\ref{coup_br}), can generate flavor-changing ALP couplings  to the down-type quarks \cite{Izaguirre:2016dfi}.
The resulting flavor-changing ALP couplings are essentially two-loop effects as the above ALP coupling to the $W$-bosons is generated by the one-loop threshold of PQ-charged heavy particles in field theoretic ALP models. 
Combining the results of \cite{Izaguirre:2016dfi} with ours, we find that the effectively two-loop ALP couplings induced by
(\ref{aWW}) dominate over our one-loop contribution if $n_u=0$ and $\tan\beta$ is large as
\bea
\label{2-loop_tb}
\tan\beta \gtrsim 17  \times \sqrt{n_H} \left[\frac{3}{C_{aWW}}\right]^{\frac{1}{2}} \left[\frac{\ln (m_{H^{\pm}}/m_t)}{2}\right]^{\frac{1}{2}} \, .
\eea 
We also remark that any new physics effect which contributes to the ALP-Higgs derivative coupling as in \cite{Bauer:2017nlg} can have an important consequence on flavor violating ALP couplings to the down-type quarks as can be seen from (\ref{fcnc_down}).

Finally, let us comment on the flavor violation in the up-type quark sector for field theoretic ALP.
For  certain class of UV models including the type-II, type Y 2HDMs and SUSY,  the bottom Yukawa coupling is enhanced by $\tan \beta$ compared to the SM.  
One may then expect  a sizable amplitude for up-type quark FCNC process for models with $n_d\neq 0$ and large $\tan \beta$ (see (\ref{fcnc_up}) and (\ref{n_s})). 
However such scenario is constrained by the following matching condition from (\ref{eq:SpurionYukawa}):
\dis{
\frac{m_b}{v} \frac{1}{ \cos\beta} = \left(\frac{\vpq}{M_*}\right)^{n_d} \lambda_b \, .
}
Again, for a cutoff scale $M_*$ significantly higher than $\vpq$, e.g.  by one order of magnitude, the perturbativity bound  $\lambda_b \lesssim {\cal O}(1)$ requires $n_d=0$  for $\tan \beta \gtrsim 10$.
On the other hand, in order for the up-quark sector to compete with the down-quark sector, we need $\tan\beta\gtrsim 20$. 
This means that  for field theoretic ALP the up-type quark sector is less sensitive to the ALP-involving flavor violation than the down-type quark sector  over the most of the ALP parameter region provided by  sensible UV completion. 

\subsection{ String theoretic ALP}

So far, we have discussed field theoretic UV completion  in which the ALP originates from the phase of PQ-charged complex scalar fields. In such models, the spontaneous breaking of PQ symmetry should be interpreted as an IR phenomenon in the context of a proper  UV completion with the cutoff scale $M_*\gg f_a$, which then  implies $n_u=0$.  
There exists in fact a totally different, but equally attractive UV completion.
ALP might originate from higher-dimensional gauge fields  in higher-dimensional theory   with an ALP decay constant $\vpq$ which has a direct connection to the fundamental scale such as the string scale or the compactification scale  \cite{Svrcek:2006yi,Choi:2003wr,Flacke:2006ad}.
The best-motivated  example is string theoretic ALP originating from $p$-form gauge field  \cite{Svrcek:2006yi} as
\bea
C_{[m_1 m_2 .. m_p]} = \sum_\alpha a_\alpha(x)\omega^\alpha_{[m_1 m_2 .. m_p]},
\eea
where $\omega^\alpha$ are harmonic $p$-form on the compact  internal space. 
Typically such ALP arises in SUSY-preserving compactification  with a modulus partner $\tau_\alpha$ describing the volume of $p$-cycle dual to $\omega^\alpha$, and forms a chiral superfield as 
\bea
T_\alpha =\frac{\tau_\alpha + ia_\alpha}{\sqrt{2}} \, ,
\eea
where we omitted the fermionic and auxiliary $F$-components.
The effective theory just below the compactification scale is described by 4D $N=1$ supergravity model with a K\"ahler potential
\bea
\label{kahler0}
K = K_0(T_\alpha+T^*_\alpha) + Z_{IJ}(T_\alpha+T^*_\alpha)\Phi_I^*\Phi_J,
\eea
where $\Phi_I$ denote the gauge-charged chiral matter superfields.
The effective theory is controlled by approximate non-linear PQ symmetries under which
\bea
a_\alpha\rightarrow a_\alpha +\mbox{constant}, 
\eea
which are the low energy remnant of the higher-dimensional gauge transformation:
\bea
\delta C_{[m_1 m_2 .. m_p]} =\partial_{[m_1}\Lambda_{m_2,..,m_p]}\, .
\eea
Note that the above non-linear PQ symmetries are defined in the GKR field basis \cite{Georgi:1986df}, so that $\Phi_I$ are invariant under the PQ symmetries.

With the K\"ahler potential (\ref{kahler0}),  the ALP effective lagrangian at the scale just below the string scale  is given by
\bea
\mathcal{L}_{eff} & = & -\frac{1}{2}\big(\partial_\alpha\partial_\beta K_0\big)\partial_\mu a_\alpha \partial^\mu a_\beta -
Z_{IJ}\left(D_\mu\phi_I^* D^\mu \phi_J - i\bar\psi_I \cancel{D} \psi_J\right)
\nonumber \\
&& - \frac{ \partial_\mu a_\alpha}{\sqrt{2}} \left[ \left(\frac{\partial Z_{IJ}}{\partial T_\alpha}\right)
\phi_I^* \overset{\leftrightarrow}{i D^\mu}\phi_J + \left(\frac{\partial Z_{IJ}}{\partial T_\alpha}-\frac{Z_{IJ}}{2}\frac{\partial K_0}{\partial T_\alpha}\right)\bar\psi_I\bar\sigma^\mu \psi_J\right] ,
\eea
where we set the reduced Planck scale $M_P=1/\sqrt{8\pi G_N}=1$.
The above lagrangian can be rewritten in terms of the canonically normalized ALP and the matter fermions and sfermions
as
% i.e. $a_p, \psi_M$ and  $\phi_M$, yielding the ALP couplings
\bea
\mathcal{L}_{eff} & = & -\frac{1}{2} \partial_\mu a_p \partial^\mu a_p -D_\mu \phi_M^* D^\mu \phi_M + i\bar\psi_M\cancel{D} \psi_M
\nonumber \\
&&-\frac{\partial_\mu a_p}{\sqrt{2}} \left[ c_{pMN} \phi_M^* \overset{\leftrightarrow}{i D^\mu}\phi_N + \left(c_{pMN}-\frac{1}{2} c_p\delta_{MN}\right)\bar\psi_M\bar\sigma^\mu \psi_N\right] \, ,
\eea
where
\bea
c_{pMN}  =  \Omega^a_{\alpha p} \Omega^\Phi_{IM}\Omega^{\Phi}_{JN} \frac{\partial Z_{IJ}}{\partial T_\alpha}, \quad
c_p  =  \Omega^a_{\alpha p} \frac{\partial K_0}{\partial T_\alpha} 
\eea
for the field redefinition matrices
\bea\Omega^a_{\alpha M}\Omega^a_{\beta N} (M_P^2\partial_\alpha\partial_\beta K_0) = \delta_{MN}, \quad  \Omega^\Phi_{I M}\Omega^\Phi_{J N} Z_{IJ} = \delta_{MN}.\eea

Unless the compactification involves a large internal space volume or an exponential warp factor, the  
stringy ALP decay constant  is generically near  $M_P/8\pi^2 \sim 10^{16}$ GeV \cite{Choi:1985je,Svrcek:2006yi}.
In such case, the flavor-changing ALP interactions would be too weak to be phenomenologically relevant.
On the other hand, in models with a large internal volume or warp factor, the resulting ALP scale can be  lower than $M_P/8\pi^2$ by many orders of magnitude, even might be around the TeV scale \cite{Antoniadis:1998ig,Burgess:1998px,Choi:2003wr,Flacke:2006ad,Balasubramanian:2005zx}.
In the following, we consider one such example, the stringy ALP in the large volume scenario (LVS) proposed in \cite{Balasubramanian:2005zx}.

For simplicity, we consider the minimal LVS with two ALPs and their modulus partners:
\bea T_1=\frac{\tau_1+ia_1}{\sqrt{2}}, \quad T_2=\frac{\tau_2+ia_2}{\sqrt{2}},\eea where
$\tau_1$ corresponds to the volume of big 4-cycle ${\cal C}_b$, which is connected to the bulk volume of the 6-dimensional internal space as
${\cal V}\sim \tau_1^{3/2}$, while $\tau_2$ is the volume of small 4-cycle ${\cal C}_s$ supporting a hidden non-perturbative dynamics, as well as the visible sector.
Following \cite{Balasubramanian:2005zx}, we assume $\tau_1$ is stabilized at an exponentially large value as
\bea
\label{lvs}
\frac{1}{\tau_1^{3/2}}\,\sim \, e^{-a\tau_2} , 
\eea
where $e^{-a\tau_2}$ parametrizes the strength of hidden non-perturbative dynamics with $a\tau_2 ={\cal O}\left({\pi^2}/{g_{\rm GUT}^2}\right)$, 
which competes with the stringy $\alpha^\prime$ corrections of ${\cal O}(1/\tau_1^{3/2})$ to stabilize $\tau_1$ at an  exponentially large vacuum value.

To be specific, let us consider the K\"ahler potential in the limit $\tau_1\gg \tau_2\gtrsim 1$, which is given by \cite{Balasubramanian:2005zx}
\bea
\label{kahler}
K=-3\ln(T_1+T_1^*) +\frac{(T_2+T_2^*)^{3/2}}{(T_1+T_1^*)^{3/2}} +\frac{(T_2+T_2^*)^{\omega_N}}{(T_1+T_1^*)}\Phi_N^*\Phi_N,
\eea 
where the modular weights $\omega_N$ of gauge charged matter superfields $\Phi_N$ are rational numbers. 
The holomorphic gauge kinetic function of the model for the SM gauge group $SU(3)_c\times SU(2)_L\times U(1)_Y$  takes the form:
\bea
\label{holomorphic_gauge}
{\cal F}_A = k_A T_2 \quad (A=3,2,1),\eea
where $k_A$ are rational numbers of order unity, and
the visible sector Yukawa couplings in the superpotential are given by
\bea
\label{homolorphic_yukawa}
\Delta W =\frac{1}{6}\lambda_{LMN}\Phi_L\Phi_M\Phi_N,\eea
where  $\lambda_{LMN}$ are independent of $T_i$ ($i=1,2$) due to the ALP shift symmetries. 
As we will see, the ALP $a_1$ associated with the big cycle has a decay constant near  $M_P$, while the small-cycle ALP $a_2$ can have a much lower decay constant in phenomenologically interesting range. 

Following the usual convention for ALP couplings, let us define the decay constant of the {\it canonically normalized}  $a_2$ though its coupling to the gauge fields.
For  the K\"ahler potential and the gauge kinetic function  given by (\ref{kahler}) and (\ref{holomorphic_gauge}), we find
\bea
\frac{1}{2}\partial_\mu a_2\partial^\mu a_2 -\frac{1}{4g_{A}^2}F^A_{\mu\nu}F^{A \mu\nu} -
\frac{1}{32\pi^2} \frac{a_2}{f_a}  F^A_{\mu\nu}\widetilde F^{A \mu\nu},\eea
where 
\bea
 \frac{1}{g_{A}^2}= k_A\frac{\tau_2}{\sqrt{2}},
\eea
and 
\bea
f_a = \frac{\sqrt{3}}{2\tau_2^{1/4}}\frac{1}{\tau_1^{3/4}}\frac{M_P}{8\pi^2}\,\sim\, e^{-a\tau_2/2}\frac{M_P}{8\pi^2}. \eea
Here we used $\tau_2={\cal O}(1/g_{\rm GUT}^2)$ and the large volume condition (\ref{lvs}) for the last expression of $f_a$.
 In the canonically normalized field basis, we find  also the following physical Yukawa couplings and the ALP couplings to the matter fields:
\bea
&&{\cal L}_{\rm Yukawa}=\frac{1}{2}y_{LMN} \phi_L\psi_M\psi_N, \nonumber \\
&&{\cal L}_{\rm inv}= \frac{\partial_\mu a_1}{\sqrt{6}M_P}
 \left(\phi_N^* \overset{\leftrightarrow}{iD^\mu}\phi_N -\frac{1}{2} \bar\psi_N\bar\sigma^\mu \psi_N\right) 
 - c_N\frac{\partial_\mu a_2}{f_a} \left(\phi_N^* \overset{\leftrightarrow}{iD^\mu}\phi_N + \bar\psi_N\gamma^\mu \psi_N\right), 
\eea
where
\bea
 y_{LMN}&=&\frac{\lambda_{LMN}}{(\sqrt{2}\tau_2)^{(\omega_L+\omega_M+\omega_N)/2}}, \nonumber \\
c_{N}&=& \frac{\sqrt{2}}{16\pi^2\tau_2}\omega_N.\eea
Note that the above couplings are defined at scales around the string scale which is related to the ALP scale as \cite{Balasubramanian:2005zx}
\bea
M_{\rm st}\sim \frac{M_P}{\tau_1^{3/4}}\sim 8\pi^2 f_a.
\eea
 
 Although the big-cycle ALP $a_1$ has a  too large decay constant to give any observable consequence in the laboratory experiments, the small-cycle ALP $a_2$ can have a decay constant in the phenomenologically interesting range, if $\tau_1$ has an exponentially large vacuum value as
$\tau_1^{3/4}\sim e^{a\tau_2/2}$ with $a\tau_2\gg 1$ \cite{Antoniadis:1998ig,Burgess:1998px,Balasubramanian:2005zx}. Yet, the pattern of  the couplings of $a_2$
is determined by the matter modular weights $\omega_N$. It has been noticed  in \cite{Conlon:2006tj} that these modular weights  can be determined by the behavior of the physical Yukawa couplings  under the rescaling of the  metric on the small-cycle ${\cal C}_s$, which results in  {\it flavor-universal}  $\omega_N$ in the range $[0,1]$.

Here are the values of modular weights in some interesting cases. One possible scenario (Case 1) is that matter zero modes live on the 4-cycle ${\cal C}_s$, with four dimensional triple intersections for Yukawa couplings, yielding  $$\omega_N=1/3.$$
Another possible scenario (Case 2) is that  matter zero modes are confined on two dimensional  curves in ${\cal C}_s$, with point-like triple intersection for Yukawa couplings, which gives $$\omega_N=1/2.$$
The final example (Case 3) we can consider
is that $Q, u^c$ and $H_2$ are confined on a singular point, while $H_1$ and/or $d^c$ can propagate over two or four dimensional surface in ${\cal C}_2$, which gives $$\omega_Q=\omega_{u^c}=\omega_{H_2}=0,  \quad \omega_{d^c}+\omega_{H_1}>0.$$
Note that $\tau_2={\cal O}(1/g_{\rm GUT}^2)$, and therefore contrary to the case of field theoretic ALP, all of these examples can be compatible with the perturbativity constraint $\lambda_{LMN}\lesssim {\cal O}(1)$, while giving the correct top quark Yukawa coupling $y_t={\cal O}(1)$.  
     
For the Cases 1 and 2, the model predicts that $n_u$ is nonzero as
\dis{
 n_u=c_Q+c_{u^c}+c_{H_2}=\frac{\sqrt{2}}{16\pi^2 \tau_2} \left(\omega_{Q}+\omega_{u^c}+\omega_{H_2} \right)={\cal O}\left(\frac{1}{16\pi^2}\right).} Although being the order of $10^{-2}$, a nonzero $n_u$ still can  yield  relatively strong flavor-changing ALP couplings to the down-type quarks due to the large logarithmic factor $\xi\ln(\Lambda_a/m_t)\simeq 2\ln(8\pi^2 f_a/m_t)$ (see eq. (\ref{fcnc_down})).
For the Case 3, $n_u=0$ and therefore the resulting flavor-changing ALP couplings to the down-type quarks are further suppressed by $1/\tan^2\beta$. However, in this case we have a nonzero $n_d$ as  
\dis{
n_d=c_Q+c_{d^c}+c_{H_1}=\frac{\sqrt{2}}{16\pi^2 \tau_2} \left(\omega_{Q}+\omega_{d^c}+\omega_{H_1}\right)={\cal O}\left(\frac{1}{16\pi^2}\right), } 
and then the resulting  ALP couplings to the up-type quarks might provide a meaningful constraint on the model if  $\tan\beta$ is large enough. 
In the next section, we will give a detailed analysis of the phenomenological constraints on the string theoretic ALP decay constant for the Cases 1 and 3.

%%%%%%%%%%%%%%%%%%%%%%%%%%
\section{Constraints on the ALP decay constant} \label{sec:Analysis}
%%%%%%%%%%%%%%%%%%%%%%%%%%

In this section, we examine the experimental constraints on the ALP decay constant $\vpq$ from flavor-changing processes, while
taking into account the properties of ALP inferred from the possible UV completions.
As we will see, the FCNC processes of down-type quarks  provide a dominant constraint on $\vpq$, 
because flavor-changing ALP couplings to the up-type quarks are suppressed by the relatively small bottom Yukawa coupling  as discussed in Sec. \ref{sec:EFT} and 
have weaker experimental sensitivity due to the long distance QCD effect \cite{Burdman:2001tf}.

According to Eqs. (\ref{fcnc_down}) and (\ref{n_s}),  flavor-changing ALP couplings to the down-type quarks in the fermion mass eigenbasis are given by
\bea
 \frac{\partial_\mu a}{\vpq}\,c^d_{ij}\, \bar{d}_i \gamma^\mu P_L d_j \, ,
\eea
where
\dis{ \label{cd_ij}
c^d_{ij}
\approx  \frac{m_t^2}{16\pi^2 v^2}(V_{\rm CKM})_{3i}^* (V_{\rm CKM})_{3j}
\left[\xi  n_u
 \ln\left(\frac{\Lambda_a}{m_{\rm BSM}}\right) +\, \left(n_u-\frac{n_H}{\tan^2 \beta} \right)\ln\left(\frac{m_{\rm BSM}}{m_t}\right) 
 \right]. 
}
 Here  $\Lambda_a\sim f_a$  for field theoretic ALP, while
$\Lambda_a\sim 8\pi^2 f_a$ for string theoretic ALP, $\xi = 1 \,(2)$ for  non-SUSY (SUSY) model, and the BSM scale $m_{\rm BSM}$ corresponds to  the charged Higgs boson  mass 
in 2HDMs, which is also taken to be the superpartner mass scale for SUSY models. 
These ALP couplings give rise to the rare meson decays such as $B\rightarrow K^{(*)} a$ and $K\rightarrow \pi a$.
We use the hadronic matrix elements using the light-cone QCD sum rules for the $B$ or $K$ meson transitions \cite{Batell:2009jf, Deshpande:2005mb, Marciano:1996wy}, yielding 
\bea
\Gamma\left(B\rightarrow Ka\right) &=&  \frac{m_B^3}{64\pi} \frac{\left| c^d_{sb}\right|^2}{\vpq^2}  \left(1-\frac{m_K^2}{m_B^2}\right)^2 \mathcal{F}_K^2\left(m_a^2\right) \lambda^{1/2}_{BKa}\,, \\
\Gamma\left(B\rightarrow K^{*}a\right) &=&   \frac{m_B^3}{64\pi} \frac{\left| c^d_{sb}\right|^2}{\vpq^2} \mathcal{F}_{K^*}^2\left(m_a^2\right) \lambda^{3/2}_{BK^*a}\,,\\
\Gamma\left(K^+\rightarrow \pi^+ a\right) &=&   \frac{m_K^3}{64\pi} \frac{\left| c^d_{ds}\right|^2}{\vpq^2} \left(1-\frac{m_\pi^2}{m_K^2}\right)^2 \lambda^{1/2}_{K\pi a}\,,\\
\Gamma\left(K_L\rightarrow \pi a\right) &=&   \frac{m_{K_L}^3}{64\pi} \frac{\left|\mathrm{Im}\left(c^d_{ds}\right)\right|^2}{\vpq^2} \left(1-\frac{m_\pi^2}{m_{K_L}^2}\right)^2 \lambda^{1/2}_{K_L\pi a}\,,
\eea
where 
\dis{
\lambda_{xyz} \equiv \left(1-\frac{(m_y+m_z)^2}{m_x^2}\right)\left(1-\frac{(m_y-m_z)^2}{m_x^2}\right),
}
and the form factors for the $B$ meson transition are given by \cite{Ball:2004ye, Ball:2004rg}
\bea
\mathcal{F}_K \left(m_a^2\right) & = & \frac{0.33}{1-m_a^2/(38 {\rm GeV^2})}\,,\\
\mathcal{F}_{K^*} \left(m_a^2\right) & = & \frac{1.35}{1-m_a^2/(28 {\rm GeV^2})} - \frac{0.98}{1-m_a^2/(37 {\rm GeV^2})} \, ,
\eea
while the form factors for the $K$ meson transition are taken to be unity.

We also have flavor-changing ALP couplings to the up-type quarks,
\bea
 \frac{\partial_\mu a}{\vpq}\,c^u_{ij}\, \bar{u}_i \gamma^\mu P_L u_j \, ,
\eea
where the coupling coefficients $c^u_{ij}$ are given in Eq.~\eqref{fcnc_up} and (\ref{n_s}):
\bea
c^u_{ij}
&\approx&  -\frac{m_b^2}{16\pi^2 v^2}  (V_{\rm CKM})_{i 3} (V_{\rm CKM})_{j 3}^* 
 \times \nonumber \\ 
 && \left[\,\xi \, n_d \, \left\{ \begin{array}{l}
\tan^2 \beta \\
1
\end{array}\right\}
 \ln\left(\frac{\Lambda_a}{m_{\rm BSM}}\right)\right. \left.+\, \left(n_d-n_H \left\{ \begin{array}{l}
1 \\
-1/\tan^2 \beta
\end{array}\right\}\right)\ln\left(\frac{m_{\rm BSM}}{m_W}\right) 
 \right], \nonumber 
 \eea
where the upper entry of the column is for the 2HDMs with $H_d = H_1$, while the lower entry corresponds to the other models with $H_d = i \sigma_2 H_2^*$. 
The most stringent constraint on the above  ALP couplings comes from the rare charm meson decay  $D^+ \rightarrow \pi^+ + a$ whose width is given by \cite{Fajfer:2015mia}
\bea
\Gamma\left(D^+\rightarrow \pi^+ a\right) =   \frac{m_{D^+}^3}{64\pi} \frac{\left| c^u_{cu}\right|^2}{\vpq^2} \left(1-\frac{m_{\pi^+}^2}{m_{D^+}^2}\right)^2 \lambda^{1/2}_{D^+\pi^+ a}\mathcal{F}_{D^+}^2 \left(m_a^2\right), 
\eea
where
\bea
\mathcal{F}_{D^+} \left(m_a^2\right)  =  \frac{0.67}{1-m_a^2/(4.58 {\rm GeV^2})}  .
\eea

The ALP produced by the rare meson decays subsequently decays into lighter SM particles
with the branching ratio determined by  the flavor-conserving ALP couplings.
In Appendix \ref{sec:LowEnergyEffectiveALPCouplings}, we provide a summary of the low energy ALP couplings  relevant for the ALP decays. 
Given the rare meson decay width to the final state involving ALP, and also the subsequent ALP decay branching ratios,  one can predict an excess in each specific rare meson decay channel over 
the background\footnote{Although our ALP model involves  BSM physics at scales above $m_{\rm BSM}$, we assume that the BSM scale is high enough, e.g. heavier than 1 TeV, so that the background event rates are essentially same as those for the SM.}.   
For instance, if the ALP decays mainly into leptons as $a \rightarrow e^+ e^-$ or $\mu^+ \mu^-$, the experimental upper bound on the branching fraction of the leptonic rare meson decay $B^+\rightarrow K^+\, l^+ l^-$  puts an upper limit on the rare meson decay width $B^+ \rightarrow  K^+ a$ times the branching ratio $a \rightarrow l^+l^-$, providing a lower bound on the ALP decay constant $\vpq$ for given values of the other model parameters.
If the ALP decay width is so small that the ALP escapes the detector before it decays,
the event will be identified as an invisible decay, constrained by the channel $K^+ \rightarrow \pi^+ + inv$, for example.
In Appendix \ref{sec:exp_constraints}, we provide a short description of the various  experimental channels which are relevant for our study. 

%%%%%%%%%%%% FIGURE %%%%%%%%%%%%%
\begin{figure}
\begin{center}
\includegraphics[width=0.47\textwidth]{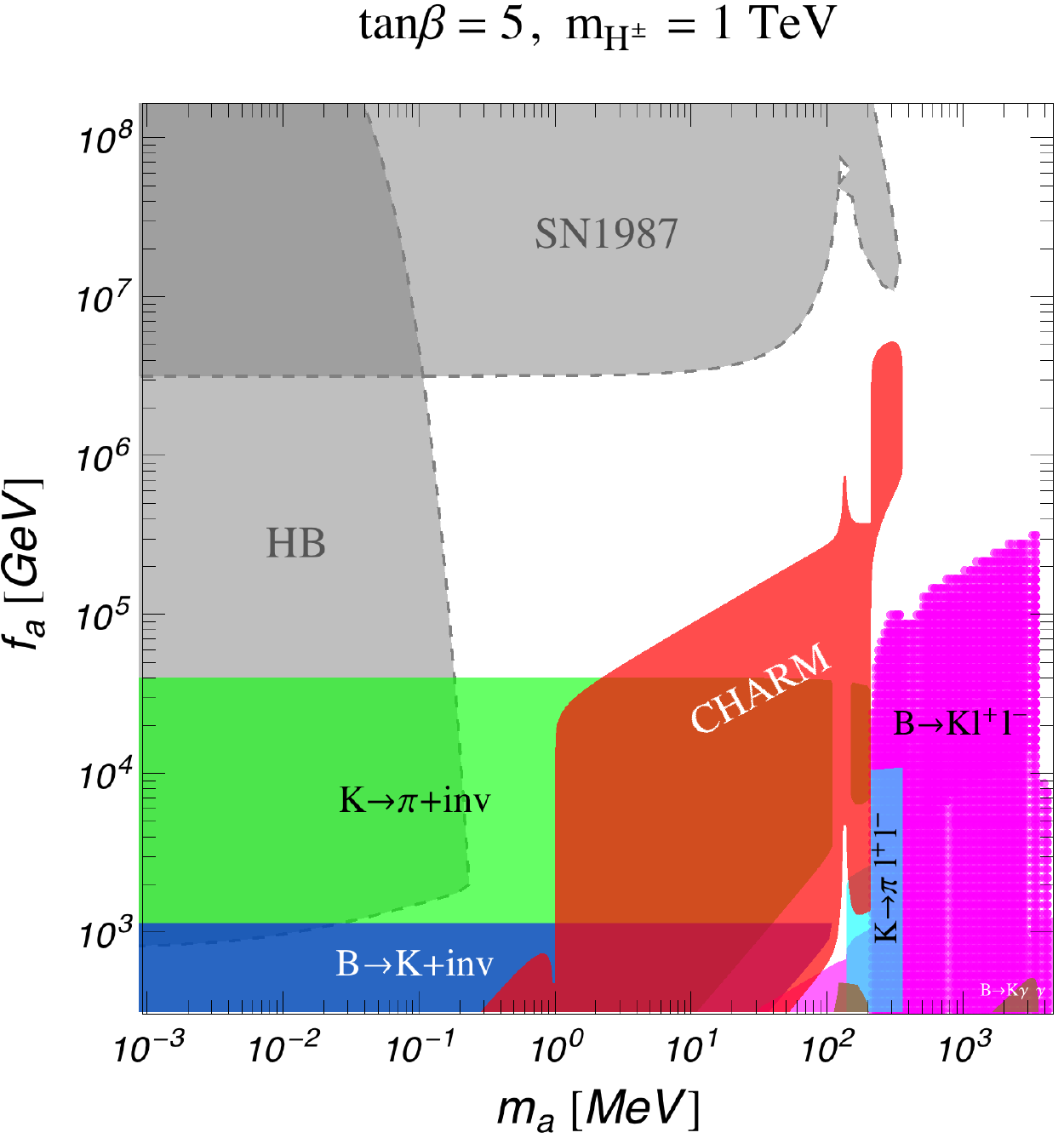}
\hspace{0.5cm}
\includegraphics[width=0.47\textwidth]{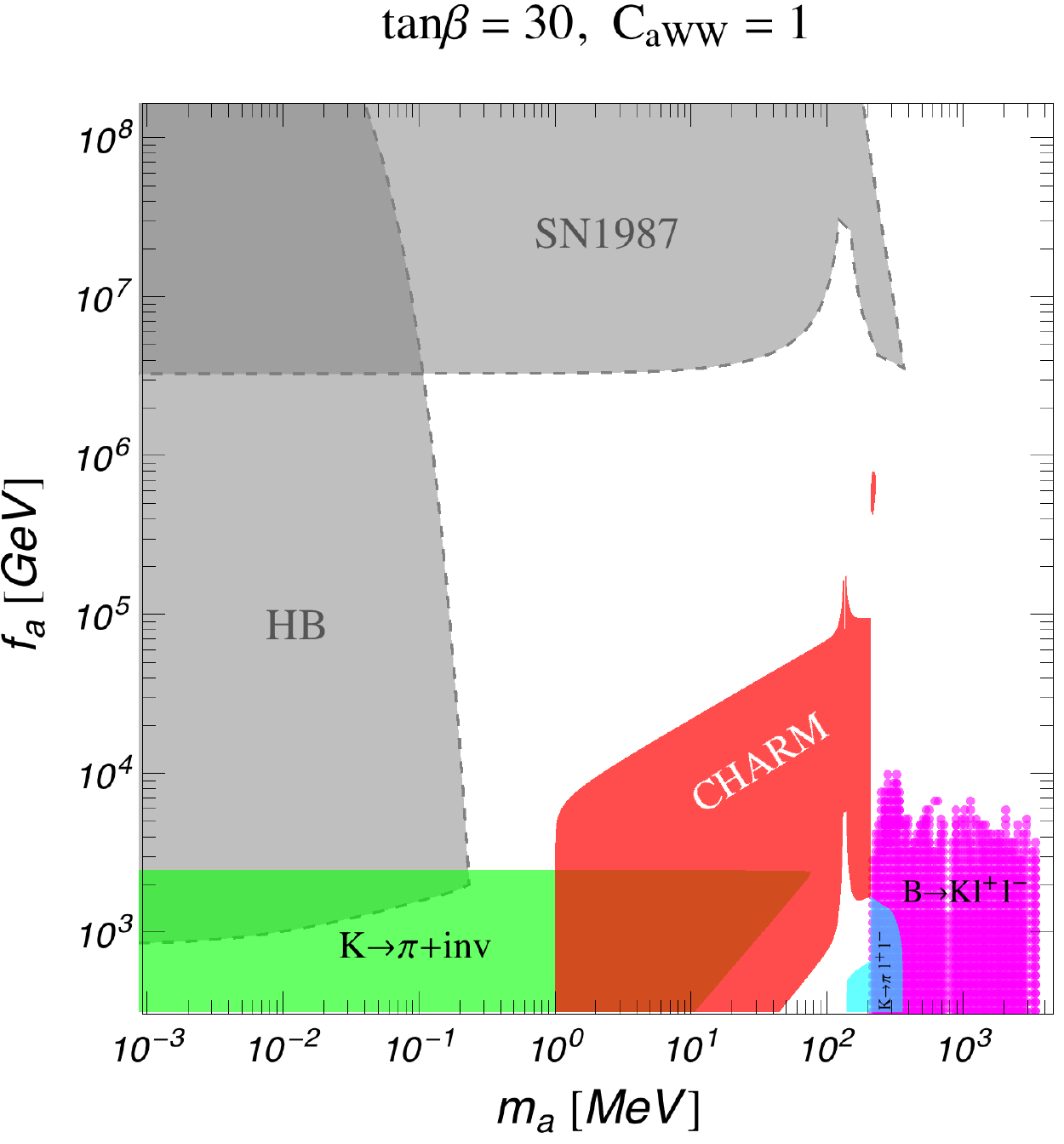}
\caption{Parameter region excluded by the FCNC constraints for field theoretic ALP ($n_u=0$ and $\Lambda_a\sim f_a$). The left panel is for the case with a moderate $\tan\beta = 5$, in which the one-loop induced ALP couplings in (\ref{cd_ij}) provide the dominant source of flavor violation. The right panel is for the case with a larger $\tan\beta=30$ satisfying (\ref{2-loop_tb}), in which
 the effective two-loop effects associated with the ALP coupling (\ref{aWW})  to the $W$-bosons, which was discussed in \cite{Izaguirre:2016dfi}, provide the dominant constraints.
 Here we consider 
the type II 2HDM with $m_{H^\pm}=1$ TeV as a benchmark model.
The results do not change much for other type of 2HDMs and SUSY models.
%{\color{red} Those plots can be regarded to be showing limiting sensitivities for sensible values of the model parameters.}
Gray parts correspond to the parameter region excluded by the conventional astrophysical considerations (SN1987 + Red giant evolution).
}
\label{fig:F-ALP}
\end{center}
\end{figure}
%%%%%%%%%%%%%%%%%%%%%%%%%%%%%

In Fig. \ref{fig:F-ALP}, we show the excluded range of the ALP decay constant $f_a$ in terms of the ALP mass $m_a$
for field theoretic ALP with  sensible UV completion, which has $n_u=0$ as discussed   in the previous section.
Although this is about a specific benchmark model, i.e. non-supersymmetric ALP model with the type II 2HD Yukawa sector, similar results are obtained also for other type of 2HD models  or SUSY models. 
The left panel corresponds to the case with $n_H=1$ and a moderate value of $\tan\beta$, in which the one-loop induced flavor-violating ALP couplings in (\ref{cd_ij})  provide the dominant source of constraints. 
The plot shows that the bound is more than an order of magnitude weaker than the results of \cite{Dolan:2014ska}. Only for ALP mass above the two muon threshold $m_a > 2m_\mu$, the bound is similar to the previous results found in  \cite{Batell:2009jf, Dolan:2014ska}, since the experimental upper limits on ${\rm Br}\left(B \rightarrow K + a (\mu^+ \mu^-)\right)$ have been significantly improved recently \cite{Aaij:2015tna, Aaij:2016qsm}.
This overall weaker bound is  due to that the condition of sensible UV completion requires $n_u=0$, and as a result the radiative correction to generate flavor-changing ALP couplings starts to operate from the 
BSM scale, which is the charged Higgs mass in our benchmark example, with a suppression by $1/\tan^2\beta$ (see Eq. (\ref{cd_ij})). 
Note that yet a sizable fraction of the parameter space for $m_a\gtrsim {\cal O}(0.1)$ MeV, which would be allowed by astrophysical  constraints, is excluded by the FCNC constraints on the radiatively generated flavor-changing ALP couplings.
If we take an even lower value of $\tan\beta$ around 1, the overall flavor constraints get severer by an order of magnitude, 
 approaching to the previous results in \cite{Dolan:2014ska} except the mass region $m_a > 2m_\mu$ where the experimental sensitivity has been upgraded. 
 This limit corresponds to the strongest flavor bound on field theoretic ALP. 
However, since such a small $\tan \beta$ would have a problem with the perturbativity bound on the top Yukawa coupling, theoretically more sensible bound is expected to be weaker
being similar to the left panel of Fig. \ref{fig:F-ALP} within order one uncertainty.

For  large $\tan \beta$ satisfying the condition (\ref{2-loop_tb}), or for the case with $n_H=0$, the effective two-loop  contribution associated with the ALP coupling (\ref{aWW})  to the $W$-bosons becomes dominant over the one-loop contribution of (\ref{cd_ij}). 
The flavor constraints in such situation  were discussed in \cite{Izaguirre:2016dfi} under the assumption that ALP does not have a tree level coupling to the charged leptons, so decays mostly into photons, which would be the case  for the KSVZ-type ALP model \cite{Kim:2008hd}.
Here  we are concerned with the DFSZ-type ALP having nonzero tree level coupling $c_\psi$ to the SM fermions, but with $n_u=c_Q+c_{u^c}+c_{H_2}=0$ for field theoretic ALP models.   As a result, in our case the ALP decays mainly  into  lepton pair, and we depict the resulting constraints in the right panel of Fig. \ref{fig:F-ALP}. 
We see that still a sizable fraction of parameter space allowed by other constraints is excluded by the flavor constraints.
Notice that this corresponds to the weakest flavor bound for the (DFSZ-type) field theoretic ALP with non-vanishing $C_{aWW}$ coupling (\ref{aWW}).
Since it is dominated by the effective two-loop contribution, it does not depend on $\tan \beta$ as long as $\tan \beta$ is large enough to satisfy the condition (\ref{2-loop_tb}). 
If $C_{aWW}=0$, the bound can be even weaker dominated by the one-loop contribution suppressed by $1/\tan^2 \beta$. In this case, we find that the lower bound on $\vpq$ becomes around TeV scale if $\tan \beta > 60$ 
for $m_a < $ 1 MeV or $m_a \gtrsim$ 100 MeV. For 1 MeV $ < m_a \lesssim $ 100 MeV, the dominant constraint comes from the CHARM beam dump experiment, which shows
a rather insensitive dependence on $\tan \beta$. For this region, the resultant lower bound on $\vpq$ is larger than 10 TeV unless $\tan \beta \gtrsim$ 100. 

%%%%%%%%%%%% FIGURE %%%%%%%%%%%%%
\begin{figure}
\begin{center}
\includegraphics[width=0.47\textwidth]{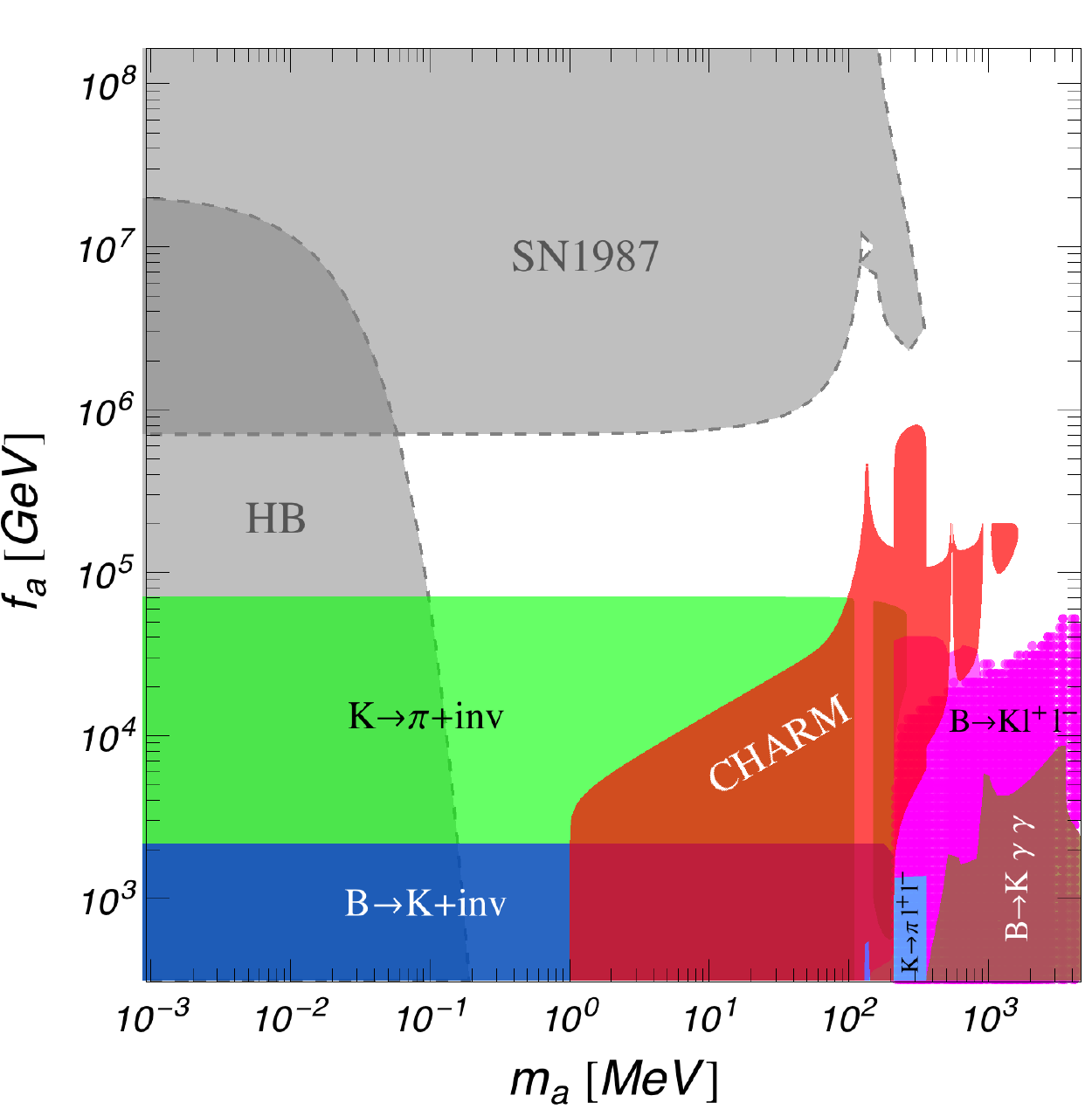}
\caption{Excluded parameter region for string theoretic ALP in the LVS scenario with universal modular weights $ \,(n_u=n_d=1/16\pi^2,\, n_H=1/24\pi^2, \, \Lambda_a\sim 8\pi^2 f_a)$.
Gray parts correspond to the parameter region excluded by the conventional astrophysical considerations (SN1987 + Red giant evolution). 
}
\label{fig:S-ALP}
\end{center}
\end{figure}
%%%%%%%%%%%%%%%%%%%%%%%%%%%%%

In Fig. \ref{fig:S-ALP}, we show the excluded parameter region for string theoretic ALP in the LVS scenario.
In the plot, we examine the case of  universal modular weights $\omega_N=1/3$ with $\tau_2 = \sqrt{2}$,
giving
\bea
n_u&=&c_Q+c_{u^c}+c_{H_2}=\frac{\sqrt{2}}{16\pi^2 \tau_2} \left(\omega_{Q}+\omega_{u^c}+\omega_{H_2} \right)=\frac{1}{16\pi^2}\nonumber \\
n_d &=&c_Q+c_{d^c}+c_{H_1}=\frac{\sqrt{2}}{16\pi^2 \tau_2} \left(\omega_{Q}+\omega_{u^c}+\omega_{H_2} \right)=\frac{1}{16\pi^2},\nonumber \\
n_H &=&c_{H_1}+c_{H_2}=\frac{\sqrt{2}}{16\pi^2 \tau_2} \left(\omega_{H_1}+\omega_{H_2} \right)=\frac{1}{24\pi^2}.\nonumber \eea
This is similar to the Yukawa-like coupling ansatz of \cite{Batell:2009jf, Dolan:2014ska}, but with additional suppression factor of $1/16\pi^2$, which is due to our convention\footnote{Note that in this convention, $c_\psi = {\cal O}(1)$ for the DFSZ-type ALP, $c_\psi = {\cal O}(\ln(f_a/\mu)/(16\pi^2)^2)$ for the KSVZ-type ALP, and $c_\psi={\cal O}(1/16\pi^2)$ for string theoretic ALP.}  to define $f_a$ in terms of the ALP couplings to gauge fields:
$$
 \frac{a}{f_a}\frac{g_A^2}{32\pi^2}F^A_{\mu\nu}\tilde F^{A\mu\nu}.
$$ However the resultant bound on $\vpq$ turns out to be only an order of magnitude weaker than the results of \cite{Batell:2009jf, Dolan:2014ska}, rather than  two orders of magnitude expected from the factor $1/16\pi^2$.
This is mostly due to the logarithmic 
factor $\xi\ln(\Lambda_a/m_t)\simeq 2\ln(8\pi^2 f_a/m_t)$ for the down-type quark flavor violation with non-zero $n_u$ as can be seen in (\ref{cd_ij}), which provides nearly an order of magnitude enhancement in our case. Note that in \cite{Batell:2009jf, Dolan:2014ska} $\Lambda_a$ is taken to be around 1 TeV, and as a result the corresponding logarithmic factor is of order unity. 
Since the major constraint  comes from the flavor-changing ALP couplings to the down-type quarks induced by  non-zero $n_u$, the bound on $\vpq$ does not depend on $\tan \beta$ and the charged Higgs mass $m_{H^\pm}$. 

In case that the matter modular weights give $n_u =0$ and $n_d\neq 0$, e.g. the Case 3 described in the previous section, the flavor constraints from the up-type quark sector might be important  if  $\tan \beta$ is large enough.
We examined this issue also, and find that for ALP mass $m_a> 100$ MeV,  the  flavor constraints from rare charm decay (with leptonic decay channel)  provide a stronger bound on $f_a$   than the down-quark sector  {\it only for} a very large  $\tan \beta >$ 70, which would constrain the ALP decay constant as $\vpq \gtrsim 1$ TeV. For smaller  $\tan \beta$, it turns out that the effective two loop flavor violation in the down-quark sector \cite{Izaguirre:2016dfi} arising from the ALP coupling (\ref{aWW}) to the W-bosons provides a stronger constraint than the up-quark sector.
However it should be noted that
 for a stringy ALP with $C_{aWW}=0$,
the up-type quark sector can be the dominant source of flavor constraints once 
$\tan \beta \gtrsim $ 20.

%%%%%%%%%%%%%%%%%%%%%%%%%%
\section{Conclusion} \label{sec:conc}
%%%%%%%%%%%%%%%%%%%%%%%%%%

In this paper, we examined  the radiatively induced flavor-changing  ALP couplings in the context of manifestly gauge-invariant  effective lagrangian, while taking into account the UV origin of the relevant bare ALP couplings. We focus on the minimal   scenario that ALP has only flavor-conserving couplings at tree level, and the dominant flavor-violating couplings are induced at one loop order due to the SM Yukawa couplings. 
As for the UV origin of ALP, we consider two possibilities : (i) field theoretic ALP originating from the phase degrees of PQ charged complex scalar fields in a UV theory with linearly realized PQ symmetry, and (ii) string theoretic ALP originating from higher dimensional $p$-form gauge fields in compactified string theory with relatively low  string scale.

For field theoretic ALP, the bare ALP parameter
 $n_u=c_Q+c_{u^c}+c_{H_2}$, which is responsible for radiative generation of the most dangerous flavor-changing ALP couplings,  is required to be vanishing in order for the underlying UV theory to admit the top quark Yukawa coupling of order unity. As can be noticed easily from the expression (\ref{cd_ij}),  this results in a 
suppression of the flavor-changing ALP couplings to down-type quarks, which is particularly efficient in the large $\tan\beta$ limit. Then, depending upon the value of $\tan\beta$, the experimental lower bound on $\vpq$ for field theoretic ALP  can be significantly  relieved  compared to the 
previous estimation  \cite{Batell:2009jf, Dolan:2014ska}, which was  based on the simple ansatz for ALP couplings that would not be realized in a sensible UV theory.

We examined also the flavor constraints on string theoretic  ALP in large volume scenario of string compactification  \cite{Balasubramanian:2005zx}, in which some of the ALPs can have a low decay constant in phenomenologically interesting range. One of the distinctive features of such string theoretic ALP is that $c_\psi ={\cal O}(1/16\pi^2)$ for  $f_a$  defined through the ALP couplings to gauge fields under the assumption  $C_A={\cal O}(1)$.
(See Eqs. (\ref{coup_sm}) and (\ref{coup_br}) for our notations.)
Note that $c_\psi={\cal O}(1)$ for DFSZ-type field theoretic ALP in the same convention \cite{Kim:2008hd}.  
Even with  $c_\psi={\cal O}(10^{-2})$, the resulting flavor constraints can be stronger than those for field theoretic ALP, in particular when $\tan\beta\gg 1$. This is
because  $n_u$ for string theoretic ALP is generically non-vanishing, although small as ${\cal O}(10^{-2})$, and therefore the flavor-violating radiative
corrections are enhanced by the large logarithmic factor $\ln(\Lambda_a/m_t)\sim \ln(8\pi^2 f_a/m_t)$ {\it without} a suppression by $1/\tan^2\beta$.

%%%%%%%%%%%%%%%%%%%%%%%%%%
\section{Acknowledgement}
%%%%%%%%%%%%%%%%%%%%%%%%%%
We thank Michael Williams and Marcin Jakub Chrzaszcz of the LHCb collaboration for informing us
the recent development on $B\rightarrow K \,a (\mu^+ \mu^-)$ searches.
This work was supported by IBS under the project code, IBS-R018-D1 [KC and SY], and by the German Science Foundation (DFG) within the SFB-Transregio TR33 ``The Dark Universe" [SHI].
The work of CBP has received funding from the European Union’s Horizon 2020 research and innovation programme under the Marie Sklodowska-Curie grant agreement No 690575.

\appendix

%%%%%%%%%%%%%%%%%%%%%%%%%%
\section{Low energy effective ALP couplings} \label{sec:LowEnergyEffectiveALPCouplings}
%%%%%%%%%%%%%%%%%%%%%%%%%%

In this appendix, we briefly summarize the flavor-conserving low energy couplings  which are relevant for the decays of axion-like particles (ALPs) which were produced by flavor-changing rare meson decays. General flavor and CP conserving effective interactions of an ALP with the SM particles at scales
just above the weak scale are given by
\bea \label{EFT_SM_UV}
&& \frac{\partial_\mu a}{\vpq}\left( c_Q \bar{Q}\,\gamma^\mu Q + c_{u^c}\, \bar{u}_R^c\gamma^\mu u_R^c + c_{d^c}\, \bar{d}_R^c\gamma^\mu d_R^c + c_L\, \bar{L}\gamma^\mu L + c_{e^c}\, \bar{e}^c_R\gamma^\mu e_R^c  + c_{H}\, H^\dagger \overset{\leftrightarrow}{i D^\mu} H\right)\nonumber \\
&&-\frac{a}{\vpq}\left( C_{agg}\,\frac{g_3^2}{32\pi^2}G\widetilde{G} + C_{aWW}\,\frac{g_2^2}{32\pi^2}W\widetilde{W} + C_{aBB}\,\frac{g_1^2}{32\pi^2}B\widetilde{B}\right) -\frac{1}{2}\hat{m}_a^2 a^2 + ...,
\eea
where $\hat{m}_a$ denotes the ALP mass {\it not} including the contribution from the ALP coupling to the gluon anomaly $G\tilde G$.
 After the electroweak symmetry breaking, the ALP-Higgs coupling induces a mixing between the ALP and $Z$-boson. Integrating out the $Z$-boson then gives a threshold correction to the ALP-fermion couplings at scales below the $Z$-boson mass.  This threshold correction can be taken into account by making the following ALP-dependent $U(1)$ rotation \cite{Georgi:1986df},
\dis{
H & \rightarrow  \exp \left(i\,c_H\,\frac{a}{\vpq}\right) H, \\
\psi & \rightarrow  \exp \left(i  \,2\,c_H\, Y_\psi \frac{a}{\vpq}\right) \psi \, ,
}
where
$Y_\psi$ denotes  the $U(1)_Y$ hypercharge of the fermion $\psi$. This hypercharge-proportional rotation enables us to
remove the ALP-Higgs coupling without affecting the ALP couplings to gauge bosons.
Then, after integrating out the massive weak gauge bosons and the top quark,  
one finds the relevant ALP interactions given by
\dis{
\frac{\partial_\mu a}{\vpq}\left[  \sum_{u_i = u, \,c} A_u \,\bar{u_i}\gamma^\mu\gamma^5 u_i + \sum_{d_i = d, \,s,\, b}  A_d \,\bar{d_i}\gamma^\mu \gamma^5 d_i  + \sum_{l = e, \,\mu, \,\tau} A_l\,\bar{l}\gamma^\mu  \gamma^5 l\right] \\ - \frac{a}{\vpq}\left( C_{agg}\frac{g_3^2}{32\pi^2}G\widetilde{G} + C_{a\gamma\gamma}\frac{e^2}{32\pi^2}F\widetilde{F}\right),
}
where now all fermions are written as the Dirac fermions, and
\dis{ 
A_u &= -\frac{1}{2}(c_Q + c_{u^c} + c_H), \\
A_d &= -\frac{1}{2}(c_Q + c_{d^c} - c_H), \\
A_l &= -\frac{1}{2}(c_L + c_{e^c} - c_H),\\
C_{a \gamma \gamma}& = C_{a WW} + C_{a BB}.
}

At the lower scale below the charm quark mass, but above the QCD scale $\Lambda_{\rm QCD}$, the relevant ALP interactions  are further reduced to 
\dis{
\frac{\partial_\mu a}{\vpq} \left[\bar{q} \gamma^\mu X_q \gamma^5 q + \bar{l} \gamma^\mu A_l  \gamma^5  l\right] - \frac{a}{\vpq}\left( C_{agg}\frac{g_3^2}{32\pi^2}G\widetilde{G} + C_{a\gamma\gamma}\frac{e^2}{32\pi^2}F\widetilde{F} \right),
}
where $q \equiv (u,d,s)$, $l \equiv (e,\mu)$, and $X_q \equiv {\rm diag}(A_u, A_d, A_d)$. 
Below the QCD scale, one should apply the chiral perturbation theory to describe the ALP interactions with mesons and baryons.  
For convenience, we first eliminate the term $a\, G\widetilde{G}$ by the following quark field rotation\footnote{For heavy ALP with a mass around the $\eta^\prime$ meson mass ($\sim 1$ GeV), this chiral rotation is no longer more convenient for calculation since the mixing between the ALP and $\eta^\prime$ becomes important. Nevertheless, we keep this approach, while keeping all kinetic or mass mixing terms in the following calculation, which would guarantee that the final results are independent of the used field basis.},
\bea
q \, \rightarrow \, \exp\left[i\frac{a}{\vpq}q_A\gamma^5\right]q,
\eea
where
\bea
{\rm Tr}\left[q_A\right] =\frac{C_{agg}}{2}.
\eea
Then the effective lagrangian becomes
\dis{
 &\frac{\partial_\mu a}{\vpq} \left[\bar{q} \gamma^\mu \left(X_q -q_A\right) \gamma^5  q + \bar{l} \gamma^\mu A_l \gamma^5  l\right] -  \bar{q}  M_q \exp \left[2i\frac{a}{\vpq}q_A\gamma^5 \right] q\\
& - \frac{a}{\vpq} \left(C_{a\gamma\gamma} -12\, {\rm Tr}\left[q_AQ_E^2\right] \right)\frac{e^2}{32\pi^2}F\widetilde{F},
}
where $M_q$ denotes the light quark mass matrix.
According to the chiral perturbation theory, the above ALP-quark couplings are matched to
\dis{
& \frac{\partial_\mu a}{\vpq}\sum_{b} j^{\mu}_{Ab} {\rm Tr}\left[\lambda_b \left( X_q - q_A\right) \right]\\
& +  \frac{1}{2}f_\pi^2\mu_\pi \left(- i\frac{a}{\vpq}{\rm Tr}\left[ \left\{ M_q,q_A \right\} \Sigma\right] -\frac{1}{2}\left(\frac{a}{\vpq}\right)^2{\rm Tr}\left[  \left\{ \left\{ M_q ,q_A \right\},q_A\right\}\Sigma\right] +{\rm h.c}+... \right),
}
where $\mu_\pi \equiv m_{\pi_0}^2 / (m_u + m_d)$ and $\lambda_a$ ($a=1,2,..,8$) are the Gell-Mann matrices. Here
the $U(3)$-valued
$\Sigma = \exp i\left({\pi_a}\lambda_a/{f_\pi} + 2{\pi_9}/\sqrt{6}{f_9}  \right)$ parametrizes 
the  pseudo-scalar mesons 
$f_9 \simeq f_\pi \simeq 93 ~{\rm MeV}$,
and the axial vector currents of meson fields are given by
\bea
&& j^\mu_{Ab} \,=\, i\frac{1}{4}f_\pi^2 \,{\rm Tr} \left[ \lambda_{b}(1-\delta_{b9})\left( \Sigma D^\mu \Sigma^{\dagger} - \Sigma^{\dagger} D^\mu \Sigma\right)\right] + \delta_{b9} f_9\partial^\mu \pi_9 \nonumber \\
&=&  f_\pi\partial^\mu \pi_{b} (1- \delta_{b9}) + \delta_{b9} f_{9}\partial^\mu \pi_9 + \delta_{b3} \frac{2}{3f_\pi}\left(\pi^0\pi^-\partial^\mu \pi^+ + \pi^0 \pi^+\partial^\mu \pi^- - 2\pi^+\pi^-\partial^\mu \pi^0\right)+ ...,
\nonumber \eea
where $\pi^0 \equiv \pi_3$ and $\pi^\pm \equiv (\pi_1 \mp i \pi_2)/\sqrt{2}$.

One can choose $q_A$ as
\begin{eqnarray} \label{qA}
q_A = \frac{C_{agg}}{2}\frac{M_q^{-1}}{{\rm Tr}\left[M_q^{-1}\right]},
\end{eqnarray}
and then the ALP mass mixing with the meson octet $\pi_a$ disappears. Then the mass-square matrix of $(\pi_3,\pi_8,\pi_9,a)$ in this field basis is given by
\dis{
 \mu_\pi \cdot \left(
\begin{tabular}{cccc}
$~m_u + m_d~$ & $~\frac{m_u - m_d}{\sqrt{3}}~$ & $~\frac{2 f_\pi}{ f_{9}}\frac{m_u- m_d}{\sqrt{6}}~$ & $~0~$ \\
$~\frac{m_u - m_d}{\sqrt{3}}~$ & $~\frac{m_u+m_d+4m_s}{3}~$ & $~\frac{2 f_\pi}{ f_{9}}\frac{m_u + m_d - 2m_s}{3\sqrt{2}}~$ & $~0~$ \\
$~\frac{2 f_\pi}{ f_{9}}\frac{m_u- m_d}{\sqrt{6}}~$ & $~\frac{2 f_\pi}{ f_{9}}\frac{m_u + m_d - 2m_s}{3\sqrt{2}}~$ &  $~\frac{3f_\pi^2}{f_{9}^2}x\,m_s + \frac{2 f_\pi^2}{f_{9}^2}\frac{m_u + m_d + m_s}{3}~$ & $~-\frac{\sqrt{6}f_\pi^2}{\vpq f_{9}}\frac{C_{agg}}{{\rm Tr}M_q^{-1}}~$ \\
$~0~$ & $~0~$ & $~-\frac{\sqrt{6}f_\pi^2}{\vpq f_{9}}\frac{C_{agg}}{{\rm Tr}M_q^{-1}}~$ & $~\frac{C_{agg}^2 f_\pi^2}{\vpq^2}\frac{1}{{\rm Tr}M_q^{-1}} + \frac{1}{\mu_\pi} \hat{m}_a^2~$ 
\end{tabular}
\right), \nonumber
}
where  $x \approx 1.68$ for the $\eta$-$\eta^\prime$ mixing angle $\theta_{\eta\eta'} \approx - 11.4 \degree$ \cite{Olive:2016xmw}, which is defined by
\bea
\left(
\begin{tabular}{c}
$\eta$ \\
$\eta'$
\end{tabular}
\right)
=
\left(
\begin{tabular}{cc}
$\cos\theta_{\eta\eta'}$ & $-\sin\theta_{\eta\eta'}$ \\
$\sin\theta_{\eta\eta'}$ & $\cos\theta_{\eta\eta'}$
\end{tabular}
\right)
\left(
\begin{tabular}{c}
$\pi_8$ \\
$\pi_9$
\end{tabular}
\right). \nonumber
\eea
We also have the following ALP-meson kinetic mixings   
\bea
\partial_\mu a \partial ^\mu \pi_3 \cdot \frac{f_\pi}{\vpq}\kappa_3 + \partial_\mu a \partial ^\mu \pi_8 \cdot \frac{f_\pi}{\vpq}\kappa_8 + \partial_\mu a \partial ^\mu \pi_9 \cdot \frac{f_{\eta'}}{\vpq}\kappa_9,
\eea
where
\bea
\kappa_3 & = &  A_u - A_d - \frac{C_{agg}}{2}\frac{m_u^{-1} - m_d^{-1}}{m_u^{-1} + m_d^{-1} + m_s^{-1}}, \nonumber \\
\kappa_8 & = & \frac{A_u - A_d}{\sqrt{3}} - \frac{C_{agg}}{2\sqrt{3}}\frac{m_u^{-1} + m_d^{-1} - 2 m_s^{-1}}{m_u^{-1} + m_d^{-1} + m_s^{-1}},\nonumber \\
\kappa_9 & = & \frac{2(A_u + 2A_d)}{\sqrt{6}} - \frac{C_{agg}}{\sqrt{6}}.\nonumber
\eea
After diagonalizing the kinetic and mass terms, we find the relevant low energy couplings of the canonically normalized mass eigenstate ALP are given by
\bea
 \frac{\partial_\mu a}{\vpq}\left[A_l \, \bar{l}\gamma^\mu\gamma^5 l +  \frac{C_{a\pi}}{f_\pi} \left(\pi^0\pi^-\partial^\mu \pi^+ + \pi^0 \pi^+\partial^\mu \pi^- - 2\pi^+\pi^-\partial^\mu \pi^0\right)\right]
 - \frac{e^2}{32\pi^2} \bar{C}_{a\gamma\gamma}\frac{a}{\vpq}F\widetilde{F} \nonumber
\eea
where
\bea
\bar{C}_{a\gamma\gamma} &\simeq& C_{a\gamma\gamma} -12~ {\rm Tr}\left[q_AQ_E^2\right] 
- 2\kappa_3 \frac{m_a^2}{m_{\pi}^2 -m_a^2} -1.3\kappa_\eta\frac{m_a^2}{m_{\eta}^2 -m_a^2}-2.9\kappa_{\eta'}\frac{m_a^2}{m_{\eta'}^2 -m_a^2},\nonumber \\
C_{a\pi} &=& \frac{2}{3} \mathrm{Tr}\left[\lambda_3 (X_q - q_A) \right],
\eea
for
\bea
\kappa_\eta & = & \kappa_8\cos\theta_{\eta\eta'}-\kappa_9\sin\theta_{\eta\eta'},\nonumber \\
\kappa_{\eta'} & = & \kappa_8\sin\theta_{\eta\eta'} + \kappa_9\cos\theta_{\eta\eta'}.\nonumber
\eea
and $q_A$ given by (\ref{qA}).

%%%%%%%%%%%%%%%%%%%%%%%%%%
\section{Summary of experimental constraints} \label{sec:exp_constraints}
%%%%%%%%%%%%%%%%%%%%%%%%%%

Here we describe the experimental constraints coming from the various rare meson decay channels used  in this paper. 
We are basically summarizing the results of Ref.~\cite{Dolan:2014ska} with some updates.

First, let us discuss the semi-invisible decay channels. 
If the decay length of ALP, i.e. $l_d \equiv |\overrightarrow{p_a} | /m_a \Gamma_a$, where $\overrightarrow{p_a}$ and $\Gamma_a$ denote the ALP momentum in the laboratory frame and the total decay width, respectively,  is much larger than the detector size, the ALP leaves no trace inside the detector. In such case, the event is to be interpreted as an invisible decay mode like
$B\rightarrow K \bar{\nu}\nu$ or $K\rightarrow \pi \bar{\nu}\nu$.
The rare $K$ decay modes $K \rightarrow \pi + inv$ have been measured by the E949 and E787 collaborations~\cite{Artamonov:2009sz}. The combined results at the $68\%$ confidence level (CL) give
\bea
{\rm Br}\left(K^+ \rightarrow \pi^+ + inv \right) \simeq \left\{ 
\begin{tabular}{ll}
$1.73^{+1.15}_{-1.05}  \times 10^{-10}$\, , & \quad$m_a = [0$ - $110] [150$ - $260]~{\rm MeV}$\\
$5.6 \times 10^{-8} $\, , & \quad$m_a \approx m_\pi $
\end{tabular}
\right.
\eea
where the second result for $m_a\approx m_\pi$ is from the E949 $90\%$ CL upper limit~\cite{Artamonov:2005cu} on ${\rm Br} (\pi^0 \rightarrow \bar{\nu}\nu) < 2.7 \times 10^{-7}$. Here, we take the detector size as 4 m.
In the near future, the proposed NA62~\cite{Anelli:2005ju} experiment will reach a sensitivity of ${\cal O}(10^{-12})$~\cite{Moulson:2016dul}.
For the rare $B$ invisible decays, the BaBar measurement~\cite{Lees:2013kla} gives the $90\%$ CL upper limits
\bea
{\rm Br}\left(B \rightarrow K + inv \right) &<& 3.2 \times 10^{-5},\nonumber \\
{\rm Br}\left(B \rightarrow K^* + inv \right) &<& 7.9 \times 10^{-5}
\eea
for the ALP mass $m_a=0-4700~{\rm MeV}$.

Next we discuss the leptonic decay channels where
the vertex resolution of detectors should be taken into account. If the ALP decay length is larger than the resolution, the event will be discarded. 
Therefore, when estimating an ALP branching ratio, one should multiply it by the probability that ALP decays within the resolution length in order to get an actual number of events to 
be taken by the detector.   
The decay mode of $K^{\pm} \rightarrow \pi^{\pm} + l^+l^-$ have been measured by the NA48/2~\cite{Batley:2009aa,Batley:2011zz} (with a vertex resolution  $\sim 1 {\rm cm}$), which results in
\bea
\left.
\begin{tabular}{ll}
${\rm Br}\left(K^{\pm} \rightarrow \pi^{\pm} + e^+e^- \right) = \left(3.11\pm0.12\right)  \times 10^{-7}  $ & $\quad (m_a=140-350~{\rm MeV}),$ \\
${\rm Br}\left(K^{\pm} \rightarrow \pi^{\pm} + \mu^+\mu^- \right) = \left(9.62\pm0.25\right)  \times 10^{-8}  $ & $\quad (m_a=210-350~{\rm MeV}),$
\end{tabular} 
\right.
\eea
where the ALP mass range relevant for each branching ratio is specified also.
The decay mode of $K_L \rightarrow \pi^0 + l^+l^-$ have been measured by the KTeV/E799~\cite{AlaviHarati:2000hs,AlaviHarati:2003mr} (with a vertex resolution  $\sim 0.4 {\rm cm}$) and the resulting $90\%$ CL upper limits on the branching ratios are  given as
\bea
\left.
\begin{tabular}{ll}
${\rm Br}\left(K_L \rightarrow \pi^0 + e^+e^- \right) < 2.8  \times 10^{-10}  $ & $\quad (m_a=140-350~{\rm MeV}),$ \\
${\rm Br}\left(K_L \rightarrow \pi^0 + \mu^+\mu^- \right) < 3.8  \times 10^{-10}  $ & $\quad (m_a=210-350~{\rm MeV}).$
\end{tabular} 
\right.
\eea
As for the decay mode $B \rightarrow K^{(*)} + l^+l^-$,  
the current world average combined result on the branching ratio on $B^+ \rightarrow K^+ +l^+ l^-$ is given as \cite{Olive:2016xmw}
\bea
{\rm Br}\left(B^+ \rightarrow K^+ + l^+l^- \right) = \left(4.51\pm 0.23\right)  \times 10^{-7}  \quad (m_a=220-4690~{\rm MeV}) .
\eea
This is in good agreement with the recent result on $B^+ \rightarrow K^+ +\mu^+ \mu^-$ from the LHCb experiment \cite{Aaij:2016cbx}.
We take the vertex resolution factor of the LHCb as $0.5$cm \cite{Dolan:2014ska}. Furthermore,
we use the recent analyses of the LHCb collaboration on  
 $B^0 \rightarrow K^{*0} + a\,(\mu^+ \mu^-) $ \cite{Aaij:2015tna} and $B^+ \rightarrow K^+ +a\,(\mu^+ \mu^-)$ \cite{Aaij:2016qsm}, which turn out to be able to put much stronger constraints depending on ALP mass and lifetime by up to $10^{-10}$ order of upper limit on the branching fraction. 
For the dimuon invariant mass  near the masses of $J\slash\psi$ and $\psi(2S)$, the long-distance effect from the charmonium resonances becomes dominant and normally screens a short-distance BSM contribution, so that one cannot simply use the above value to constrain the ALP physics \cite{Aaij:2015tna, Aaij:2016qsm, Aaij:2016cbx}. 
Yet, the branching ratio of $B^+\rightarrow K^+ + a$ with $a \rightarrow l^+ l^-$ should not exceed  ${\rm Br}(B^+\rightarrow K^+ + J\slash \psi \rightarrow K^+ + l^+l^-)$ and ${\rm Br}(B^+\rightarrow K^+ + \psi(2S)\rightarrow K^+ + l^+l^-)$~\cite{Olive:2016xmw},  and therefore 
\bea
\left.
\begin{tabular}{ll}
${\rm Br}\left(B^+ \rightarrow K^+ + l^+l^- \right) < 6.0  \times 10^{-5}  $ & $\quad (m_a=2950-3180~{\rm MeV})$, \\
${\rm Br}\left(B^+ \rightarrow K^+ + l^+l^- \right) < 4.9 \times 10^{-6}   $ & $\quad (m_a=3590-3770~{\rm MeV})$.
\end{tabular} 
\right.
\eea

Let us now discuss the photon decay channels.
The decay mode of $K_L \rightarrow \pi^0 + \gamma\gamma$ have been measured by the KTeV~\cite{Abouzaid:2008xm}, which results in 
\dis{
{\rm Br}\left(K_L \rightarrow \pi^0 + \gamma\gamma \right) = &\left(1.29\pm 0.03_{\rm stat} \pm 0.05_{\rm sys}\right)  \times 10^{-6}. \, \\ &(m_a=[40\text{-}100],[160\text{-}350]~{\rm MeV}) 
}
For $m_a \sim m_\pi$, the SM background from $K_L \rightarrow \pi^0\pi^0$ reduces the sensitivity.
As in the case of rare $B$ meson leptonic decays, the branching ratio of $K_L \rightarrow \pi^0+a$ with $a \rightarrow\gamma\gamma$ should not exceed  ${\rm Br}(K_L \rightarrow \pi^0\pi^0\rightarrow \pi^0 + \gamma\gamma)$~\cite{Olive:2016xmw,Alexopoulos:2004sx}, implying  
\bea
{\rm Br}\left(K_L \rightarrow \pi^0 + \gamma\gamma \right) <8.6  \times 10^{-4}  \quad (m_a\sim m_\pi).
\eea
The rare $B$ decay mode $B\rightarrow K +\gamma\gamma$ has been measured previously by the B-factories (BaBar~\cite{Aubert:2007hh} and Belle~\cite{Duh:2012ie} with a relatively large vertex resolution $\sim 30 ~{\rm cm}$), but
only for the diphoton invariant mass $m_{\gamma\gamma} \sim m_\pi$.
Since the measured branching fraction is order of $10^{-5} \sim 10^{-6}$, we choose a conservative upper limit of $10^{-6}$ for the ALP decay mode.

As for the flavor constraints coming from the up-type quarks, the rare charm meson decay can be relevant.
In spite of the long distance  QCD effect screening short distance physics, the process $D^+ \rightarrow \pi^+ \mu^+\mu^-$ with  dimuon invariant mass which is potentially sensitive to short distance BSM physics has been measured by the LHCb,  yielding the following  $90\%$ CL upper limit on the branching ratio:
\bea
{\rm Br}\left(D^+ \rightarrow \pi^+ \mu^+\mu^-\right) \simeq \left\{ 
\begin{tabular}{ll}
$2.0  \times 10^{-8}$ & \quad$(m_a = [250$ - $525]~{\rm MeV}),$\\
$2.6 \times 10^{-8} $ & \quad$(m_a = [1250$ - $1700]~{\rm MeV}),$\\
$7.3 \times 10^{-8} $ & \quad (total).
\end{tabular}
\right.
\eea

Finally, the beam dump experiment searching for long-lived light particle can also constrain the ALP FCNC processes~\cite{Dobrich:2015jyk,Bezrukov:2009yw}.
It turns out that presently the CHARM experiment~\cite{Bergsma:1985qz}
using the proton-proton beam collision
gives the most stringent constraint.
The total number of produced ALPs can be estimated by the following ratio to the pion production cross section \cite{Bezrukov:2009yw, Clarke:2013aya}: 
\bea
N_a \approx \left(2.9 \times 10^{17}\right) \cdot \frac{\sigma_a}{\sigma_\pi},
\eea
where
\dis{
\frac{\sigma_a}{\sigma_\pi} \approx 3 \cdot \left( \frac{1}{14}\,\text{Br}\left(K^+\rightarrow \pi^+ + a\right) +  \frac{1}{28}\,\text{Br}\left(K_L\rightarrow \pi^0 + a\right) +  3\cdot 10^{-8}\,\text{Br}\left(B\rightarrow X + a\right)\right).\nonumber
}
Since the detector is $35 \mathrm{m}$ long and $480 \mathrm{m}$ away from the target, the number of the signals from ALP decays is estimated as
\bea
N_d \approx N_a \cdot \text{Br}\left(a \rightarrow \gamma\gamma, ee, \mu\mu\right)\cdot \left[\exp\left(-\Gamma_a\frac{480\mathrm{m}}{\gamma}\right)-\exp\left(-\Gamma_a\frac{515\mathrm{m}}{\gamma}\right)\right] 
\eea
with $\gamma \simeq 25\,\mathrm{GeV}/m_a$.
From that there is no signal from CHARM experiment, one then finds the 90 $\%$ CL bound \cite{Clarke:2013aya}\bea
N_d < 2.3.\eea

\end{document}